\documentclass[twocolumn]{aastex63}
\usepackage[nameinlink]{cleveref}
\usepackage{ragged2e}
\usepackage[]{natbib}
\newcommand\degree{$^{\circ}$}
\newcommand\degrees\degree

\DeclareSymbolFont{UPM}{U}{eur}{m}{n}
\DeclareMathSymbol{\umu}{0}{UPM}{"16}
\let\oldumu=\umu
\renewcommand\umu{\ifmmode\oldumu\else\math{\oldumu}\fi}

\newcommand\microns \micron

\let\oldsim=\sim
\renewcommand\sim{\ifmmode\oldsim\else\math{\oldsim}\fi}
\let\oldpm=\pm
\renewcommand\pm{\ifmmode\oldpm\else\math{\oldpm}\fi}
\newcommand\by{\ifmmode\times\else\math{\times}\fi}

\newbox{\wdbox}
\renewcommand\c{\setbox\wdbox=\hbox{,}\hspace{\wd\wdbox}}
\renewcommand\i{\setbox\wdbox=\hbox{i}\hspace{\wd\wdbox}}
\newcommand\n{\hspace{0.5em}}

\newcount\timect
\newcount\hourct
\newcount\minct
\newcommand\now{\timect=\time \divide\timect by 60
         \hourct=\timect \multiply\hourct by 60
         \minct=\time \advance\minct by -\hourct
         \number\timect:\ifnum \minct < 10 0\fi\number\minct}

\catcode`@=11

\newcommand\comment[1]{}
\newcommand\commenton{\catcode`\%=14}
\newcommand\commentoff{\catcode`\%=12}

\renewcommand\math[1]{$#1$}
\newcommand\mathshifton{\catcode`\$=3}
\newcommand\mathshiftoff{\catcode`\$=12}

\comment{the backslash is necessary}

\comment{alignment tab}

\let\atab=&
\newcommand\atabon{\catcode`\&=4}
\newcommand\ataboff{\catcode`\&=12}

\let\oldmsp=\sp
\let\oldmsb=\sb
\def\sp#1{\ifmmode
           \oldmsp{#1}%
         \else\strut\raise.85ex\hbox{\scriptsize #1}\fi}
\def\sb#1{\ifmmode
           \oldmsb{#1}%
         \else\strut\raise-.54ex\hbox{\scriptsize #1}\fi}
\newbox\@sp
\newbox\@sb
\def\sbp#1#2{\ifmmode%
           \oldmsb{#1}\oldmsp{#2}%
         \else
           \setbox\@sb=\hbox{\sb{#1}}%
           \setbox\@sp=\hbox{\sp{#2}}%
           \rlap{\copy\@sb}\copy\@sp
           \ifdim \wd\@sb >\wd\@sp
             \hskip -\wd\@sp \hskip \wd\@sb
           \fi
        \fi}
\def\msp#1{\ifmmode
           \oldmsp{#1}
         \else \math{\oldmsp{#1}}\fi}
\def\msb#1{\ifmmode
           \oldmsb{#1}
         \else \math{\oldmsb{#1}}\fi}
\def\supon{\catcode`\^=7}
\def\supoff{\catcode`\^=12}
\def\subon{\catcode`\_=8}
\def\suboff{\catcode`\_=12}
\def\supsubon{\supon \subon}
\def\supsuboff{\supoff \suboff}

\newcommand\actcharon{\catcode`\~=13}
\newcommand\actcharoff{\catcode`\~=12}

\newcommand\paramon{\catcode`\#=6}
\newcommand\paramoff{\catcode`\#=12}

\comment{And now to turn us totally on and off...}

\newcommand\reservedcharson{\commenton \mathshifton \atabon \supsubon \actcharon
	\paramon}

\newcommand\reservedcharsoff{\commentoff \mathshiftoff \ataboff
	\supsuboff \actcharoff \paramoff}

\catcode`@=12
\reservedcharsoff

\reservedcharson

\comment{ Must have ONLY ONE of these... trust these macros, they work

}

\newcommand{\squishlist}{
 \begin{list}{$\bullet$}
  { \setlength{\itemsep}{0pt}
     \setlength{\parsep}{0pt}
     \setlength{\topsep}{0pt}
     \setlength{\partopsep}{0pt}
     \setlength{\leftmargin}{2.0em}
     \setlength{\labelwidth}{1.5em}
     \setlength{\labelsep}{0.5em} } }

\newcommand{\squishlisttwo}{
 \begin{list}{$\bullet$}
  { \setlength{\itemsep}{1pt}
     \setlength{\parsep}{3pt}
     \setlength{\topsep}{3pt}
     \setlength{\partopsep}{0pt}
     \setlength{\leftmargin}{2.0em}
     \setlength{\labelwidth}{1.5em}
     \setlength{\labelsep}{0.5em} } }

\newcommand{\squishend}{
  \end{list}  }

\reservedcharson
\actcharon
\usepackage[utf8]{inputenc}
\usepackage[T1]{fontenc}
\usepackage{amssymb}
\usepackage{latexsym}
\usepackage{stmaryrd}

\submitjournal{AJ}

\shorttitle{Spitzer phase curves of WASP-76\lowercase{b}}
\shortauthors{May \& Komacek, et al.}
\graphicspath{{./}{figs/}}

\begin{document}

\title{Spitzer phase curve observations and circulation models of the inflated ultra-hot Jupiter WASP-76b}

\correspondingauthor{Erin May}
\email{Erin.May@jhuapl.edu}
\author[0000-0002-2739-1465]{Erin M. May}
\altaffiliation{These authors contributed equally to this work.}
\affiliation{Johns Hopkins APL, 11100 Johns Hopkins Rd, Laurel, MD 20723, USA}
\author[0000-0002-9258-5311]{Thaddeus D. Komacek}
\altaffiliation{These authors contributed equally to this work.}
\affiliation{Department of the Geophysical Sciences, University of Chicago, Chicago, IL 60637, USA}
\author[0000-0002-7352-7941]{Kevin B. Stevenson}
\affiliation{Johns Hopkins APL, 11100 Johns Hopkins Rd, Laurel, MD 20723, USA}
\author[0000-0002-1337-9051]{Eliza M.-R. Kempton}
\affiliation{Department of Astronomy, University of Maryland, College Park, MD 20742, USA}
\author[0000-0003-4733-6532]{Jacob L. Bean}
\affiliation{Department of Astronomy \& Astrophysics, University of Chicago, 5640 S. Ellis Avenue, Chicago, IL 60637, USA}
\author[0000-0002-2110-6694]{Matej Malik}
\affiliation{Department of Astronomy, University of Maryland, College Park, MD 20742, USA}
\author{Jegug Ih}
\affiliation{Department of Astronomy, University of Maryland, College Park, MD 20742, USA}
\author[0000-0003-4241-7413]{Megan Mansfield}
\affiliation{Department of the Geophysical Sciences, University of Chicago, Chicago, IL 60637, USA}
\author[0000-0002-2454-768X]{Arjun B. Savel}
\affiliation{Department of Astronomy, University of Maryland, College Park, MD 20742, USA}
\author{Drake Deming}
\affiliation{Department of Astronomy, University of Maryland, College Park, MD 20742, USA}
\author{Jean-Michel Desert}
\affiliation{Anton Pannekoek Institute for Astronomy, University of Amsterdam, 1090 GE Amsterdam, Netherlands}
\author{Y. Katherina Feng} 
\affiliation{Department of Astronomy \& Astrophysics, University of California, Santa Cruz, CA 95064, USA}
\author[0000-0002-9843-4354]{Jonathan J. Fortney} 
\affiliation{Department of Astronomy \& Astrophysics, University of California, Santa Cruz, CA 95064, USA}
\author{Tiffany Kataria} 
\affiliation{NASA Jet Propulsion Laboratory, 4800 Oak Grove Drive, Pasadena, CA, USA}
\author{Nikole Lewis} 
\affiliation{Department of Astronomy and Carl Sagan Institute, Cornell University, Ithaca, NY 14853, USA}
\author{Caroline Morley}
\affiliation{Department of Astronomy, University of Texas at Austin, Austin, TX 78712, USA}
\author{Emily Rauscher}
\affiliation{Department of Astronomy, University of Michigan, Ann Arbor, MI 48109, USA}
\author{Adam Showman}
\altaffiliation{Deceased.}
\affiliation{Lunar and Planetary Laboratory, University of Arizona, Tucson, AZ, 85721, USA}





\begin{abstract}
The large radii of many hot Jupiters can only be matched by models that have hot interior adiabats, and recent theoretical work has shown that the interior evolution of hot Jupiters has a significant impact on their atmospheric structure. Due to its inflated radius, low gravity, and ultra-hot equilibrium temperature, WASP-76b is an ideal case study for the impact of internal evolution on observable properties. Hot interiors should most strongly affect the non-irradiated side of the planet, and thus full phase curve observations are critical to ascertain the effect of the interior on the atmospheres of hot Jupiters. In this work, we present the first Spitzer phase curve observations of WASP-76b. We find that WASP-76b has an ultra-hot day side and relatively cold nightside with brightness temperatures of $2471 \pm 27~\mathrm{K}$/$1518 \pm 61~\mathrm{K}$ at $3.6~\micron$ and $2699 \pm 32~\mathrm{K}$/$1259 \pm 44~\mathrm{K}$  at $4.5~\micron$, respectively. These results provide evidence for a dayside thermal inversion. Both channels exhibit small phase offsets of $0.68 \pm 0.48^{\circ}$ at $3.6~\micron$ and $0.67 \pm 0.2^{\circ}$ at $4.5~\mu\mathrm{m}$. We compare our observations to a suite of general circulation models that consider two end-members of interior temperature along with a broad range of frictional drag strengths. Strong frictional drag is necessary to match the small phase offsets and cold nightside temperatures observed. From our suite of cloud-free GCMs, we find that only cases with a cold interior can reproduce the cold nightsides and large phase curve amplitude at $4.5~\micron$, hinting that the hot interior adiabat of WASP-76b does not significantly impact its atmospheric dynamics or that clouds blanket its nightside. 
\end{abstract}



\section{Introduction} 
\label{sec:intro}
Ultra-hot Jupiters, gas giants with equilibrium temperatures $\gtrsim 2200~\mathrm{K}$, have recently been recognized to be a new class of exoplanets with observable properties that are distinct from cooler hot Jupiters. Notably, spectra of ultra-hot Jupiters generally show diminished absorption features \citep{Stevenson2014b,Haynes:2015,Beatty:2017aa,Evans:2017aa,Arcangeli:2018aa,Kreidberg:2018aa,Mansfield:2018aa,Baxter:2020aa,Fu:2020aa,Mikal-Evans:2020aa} due to a combination of thermal dissociation of molecules (most importantly H$_2$O) and H$^{-}$ continuum opacity \citep{Parmentier:2018aa,Lothringer:2018aa,Kitzmann:2018aa}. Recent work has found population-level evidence for dayside thermal inversions in ultra-hot Jupiters, but not hot Jupiters, from a large sample of Spitzer secondary eclipse depths \citep{Baxter:2020aa,Garhart20,Mansfield:2021aa}. 

While secondary eclipse spectra are able to probe the presence (or lack) of a thermal inversion on the planet's dayside, phase resolved observations are required to make this determination on the planet's nightside. Nightside thermal inversions on tidally locked gas giant planets are generally not expected due to the lack of stellar flux combined with the low efficiency of heat transport at the pressure levels in question. However, nightside observations from Spitzer phase curves have often been inconclusive or suggestive of atmospheric conditions not in line with these predictions from circulation models, potentially due to cloud-free models while observations imply a uniform nightside cloud coverage as identified in \cite{Beatty2019} and \cite{Keating:2019aa}. For example, \cite{Wong:2015} find higher than expected nightside emission at 4.5 $\micron$ in the atmosphere of WASP-14b, indicative of an excess absorber, potentially explained by an emitting layer of CO in the upper atmosphere, or simply an enhanced C/O ratio. 
More recently, \cite{Kreidberg:2018aa} find evidence for a thermal inversion on the dayside of the ultra-hot Jupiter WASP-103b with no inversion on the nightside. As an ultra-hot Jupiter with an inflated radius, WASP-76b is a particularly useful target for further exploring the effect of inversions in the atmospheres of this class of planets.

Novel processes play a key role in the atmospheric physics and chemistry of ultra-hot Jupiters and can significantly impact their atmospheric circulation. As hydrogen is partially dissociated on the day sides of ultra-hot Jupiters, the latent heat release of hydrogen recombination on the cool night side and intake through hydrogen dissociation on the hot day side is expected to greatly impact the phase curve amplitudes of ultra-hot Jupiters \citep{Bell:2018aa,Komacek:2018aa,Tan:2019aa,Gandhi:2020aa}. Tentative evidence of the effect of hydrogen dissociation and recombination on heat transport has been seen in recent Spitzer and TESS phase curve observations of the ultra-hot Jupiter KELT-9b \citep{Mansfield:2020aa,Wong:2019aa}.

Most recently, ground-based high-resolution spectroscopy (spectral resolution $\gtrsim 25,000$) has shown a wealth of extreme chemistry in the hot upper atmospheres of ultra-hot Jupiters through a broad range of discoveries of both atomic and ionized metal lines and hydrogen Balmer features \citep{Nugroho:2017aa,Hoeijmakers:2018aa,Jensen:2018aa,Seidel:2019aa,Cabot:2020aa,Ehrenreich:2020aa,Hoeijmakers:2020aa,Wyttenbach:2020aa,Yan:2020aa}. These high-resolution spectroscopic observations also provide a path toward directly probing the atmospheric dynamics of hot gas giants through measuring the Doppler shifts and broadening of spectral lines \citep{Kempton:2012aa,showman_2013_doppler,Kempton:2014,Louden:2015,Brogi:2015,Zhang:2017b,Flowers:2018aa,Beltz:2021aa}. 

WASP-76b has emerged as a baseline ultra-hot Jupiter due to its inflated radius of $1.85~R_\mathrm{Jup}$ and bright ($m_v = 9.5$) F7 host star \citep{West:2016aa}. A broad range of previous studies \citep{Fu:2017aa,Tsiaras:2018aa,Fisher18,Edwards:2020aa,Essen:2020aa,Fu:2020aa,Garhart20} have analyzed Spitzer and HST transmission and emission observations of WASP-76b. The emission spectrum of WASP-76b is typical of ultra-hot Jupiters, with a diminished water absorption feature due to dissociation and an emission feature due to CO at $4.5~\mu\mathrm{m}$ indicative of a thermal inversion \citep{Fu:2020aa}. This thermal inversion may be due to a combination of the strong optical absorption by metals and/or TiO and reduced longwave cooling due to water dissociation near the photosphere \citep{Lothringer:2018aa}. The strong optical opacity of metals also enhances the transit depth at optical wavelengths \citep{Lothringer:2020aa}, and the largely cloud-free atmosphere of WASP-76b as seen in both transmission and emission \citep{Fu:2020aa} provides an opportunity to constrain atmospheric elemental ratios and link the composition to formation scenarios \citep{Lothringer:2020ab,Ramirez:2020aa}. 

Ground-based high spectral resolution observations have also constrained the properties of the upper atmosphere of WASP-76b, detecting a range of atomic features with ESPRESSO \citep{Ehrenreich:2020aa,Tabernero:2020aa} and HARPS \citep{Seidel:2019aa,Zak:2019aa}. Most notably, \cite{Ehrenreich:2020aa} found evidence for iron condensation on the night side of the planet, as the iron absorption signal was asymmetric between the eastern and western limb. This detection of asymmetric iron absorption was recently confirmed with archival HARPS observations \citep{Kesseli:2021aa}. \cite{Ehrenreich:2020aa} also found evidence for a uniform day-to-night wind with a blueshift of $\approx 5.3~\mathrm{km}~\mathrm{s}^{-1}$ on both limbs after compensating for planetary rotation. Additionally, \cite{Seidel:2019aa} found that sodium lines in the upper atmosphere of WASP-76b are strongly broadened, which may be indicative of an evaporative torus of material surrounding the planet \citep{Gebek:2020aa}.

As WASP-76b is a highly inflated and low gravity planet with an ultra-hot equilibrium temperature, a variety of processes are expected to affect its atmospheric dynamics and resulting spatial pattern of emitted flux. Notably, the present-day radius of WASP-76b indicates the need for a hot interior adiabat in order to provide a sufficient central entropy to match its present-day radius \citep{Arras:2006kl,Spiegel:2013,Komacek:2017a,Thorngren:2017,Thorngren:2019aa,Sarkis:2021aa}. In this work, we will test this theoretical expectation of a hot interior through measuring its impact on the atmospheric circulation with phase curve observations. 

If WASP-76b has a hot internal adiabat, it would coincide with a high internal heat flux that may drive a strong magnetic field \citep{Christensen:2009,Yadav:2017} as has been observed for other hot Jupiters through signals of star-planet interactions \citep{Cauley:2019aa}. The combination of a strong internal magnetic field and a hot and thermally ionized atmosphere would imply that magnetic effects (e.g., Lorentz forces) affect the atmospheric circulation of WASP-76b \citep{Perna_2010_1,Menou:2012fu,batygin_2013,Rauscher_2013,Rogers:2020,Rogers:2017,Rogers:2017a}. The strong atmospheric flows driven by the large day-to-night irradiation contrast may also undergo large-scale turbulence due to fluid instabilities and/or shocks \citep{Li:2010,Heng:2012a,Menou_2012,perna_2012,Polichtchouk:2012,Fromang:2016,Koll:2017,Menou:2019aa}. As expected from the iron condensation observed by \cite{Ehrenreich:2020aa}, condensate clouds likely form on the night side and western limb of WASP-76b and radiatively feed back on the temperature structure \citep{Helling:2016,Parmentier16,Wakeford:2017,Powell:2018aa,Gao:2020aa,Roman:2020aa,Helling:2021aa}. Additionally, this non-uniform aerosol distribution can impact observable properties (especially in transmission, \citealp{Fortney05c,Kempton:2017aa,Powell:2019aa}) and lead to a dynamic interplay with the atmospheric circulation \citep{Lee:2016,Lines:2018,lines:2019,Roman:2019aa,Parmentier:2020aa,Roman:2020aa,Steinrueck:2020aa}. 

In this work, we constrain the atmospheric circulation of WASP-76b in detail utilizing the first Spitzer 3.6 and 4.5 $\micron$ phase curve observations of this planet. We conduct novel general circulation model (GCM) simulations of the atmospheric circulation of WASP-76b over a wide range of atmospheric frictional drag strength and internal temperature, and utilize these GCMs to interpret the observed Spitzer phase curve. This manuscript is organized as follows. In Section \ref{sec:obs}, we describe the data reduction of and results from our Spitzer 3.6 and 4.5 $\micron$ phase curve observations. We then describe the setup of our GCMs and present results from simulations with varying internal temperature and atmospheric drag in Section \ref{sec:GCM}. We compare these GCM simulations to our observed phase curve in Section \ref{sec:disc}, and we outline our conclusions in Section \ref{sec:conc}.
\section{Spitzer observations}
\label{sec:obs}
\subsection{Observations}
Three phase curves of WASP-76b were observed with Spitzer's InfraRed Array Camera \citep[IRAC,][]{IRAC}, two at 3.6 $\micron$ and one at 4.5 $\micron$ (Program 13038, PI: Kevin Stevenson). Each phase curve covers approximately 51 hours, including one transit and two secondary eclipses, with two-second frame times at 4.5 $\micron$ and 0.4-second frame times at 3.6 $\micron$. Table \ref{table:observations} overviews these observations.

\begin{figure*}
    \centering
    \includegraphics[width = 0.325\textwidth]{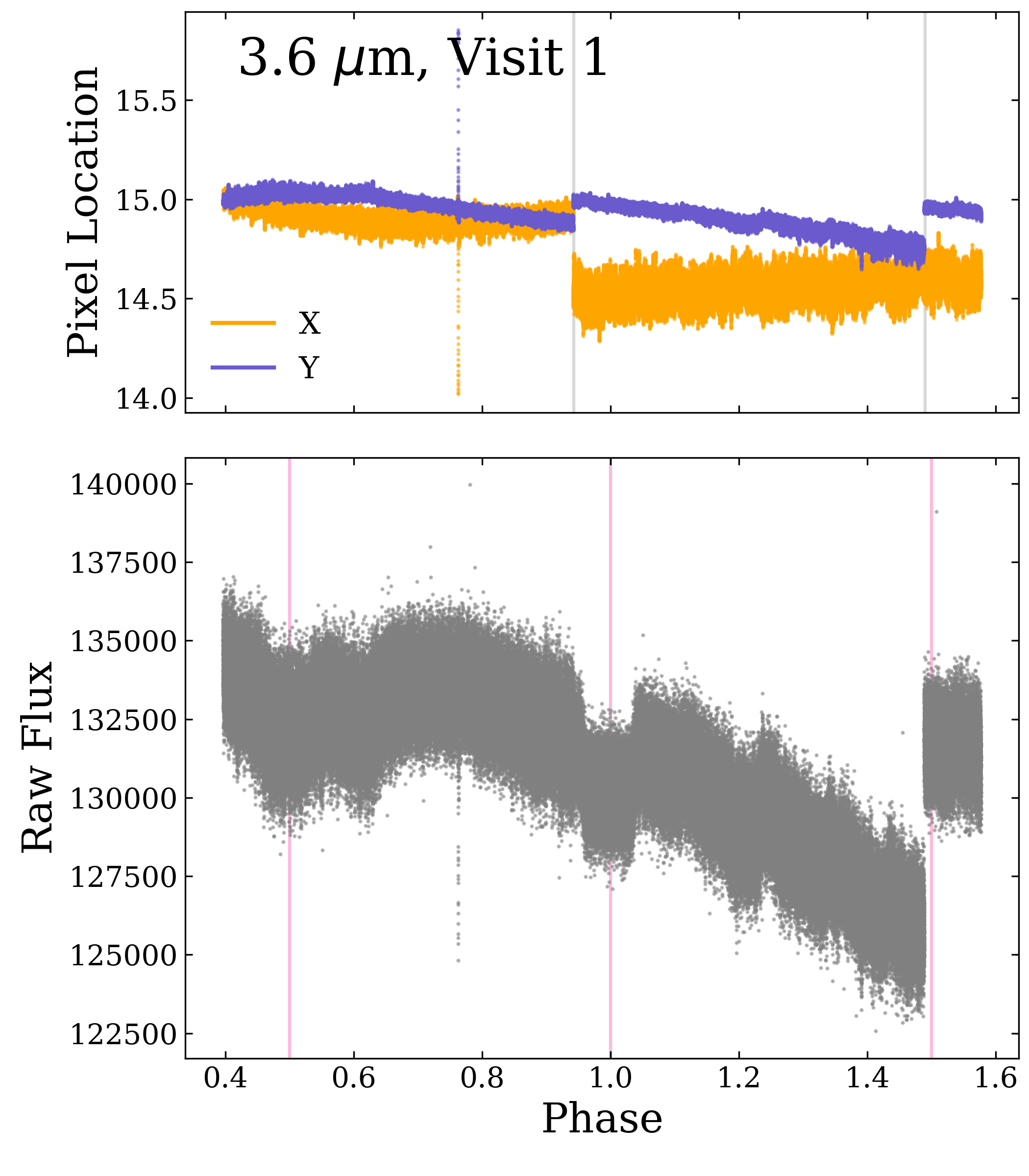}
    \includegraphics[width = 0.325\textwidth]{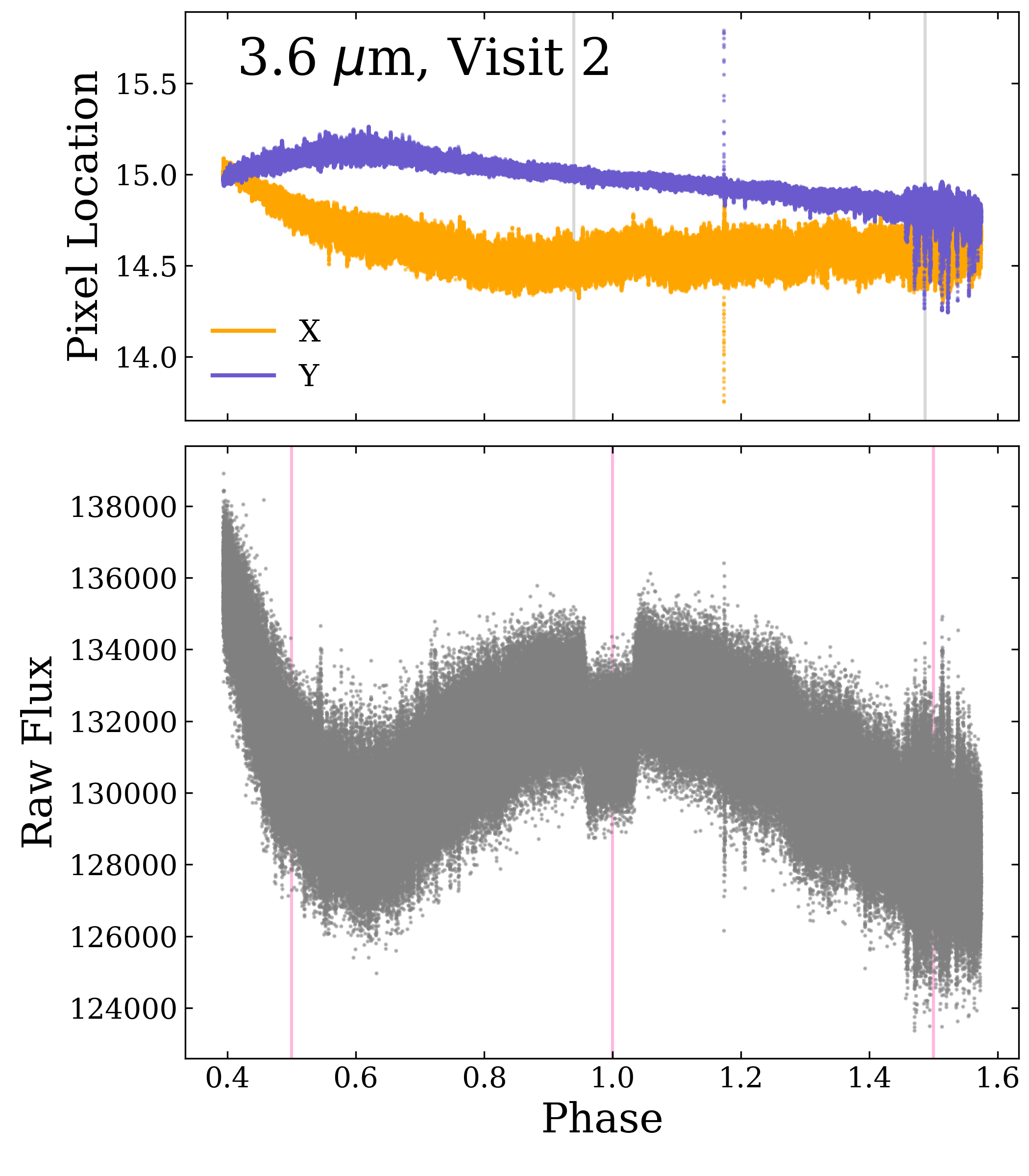}
    \includegraphics[width = 0.325\textwidth]{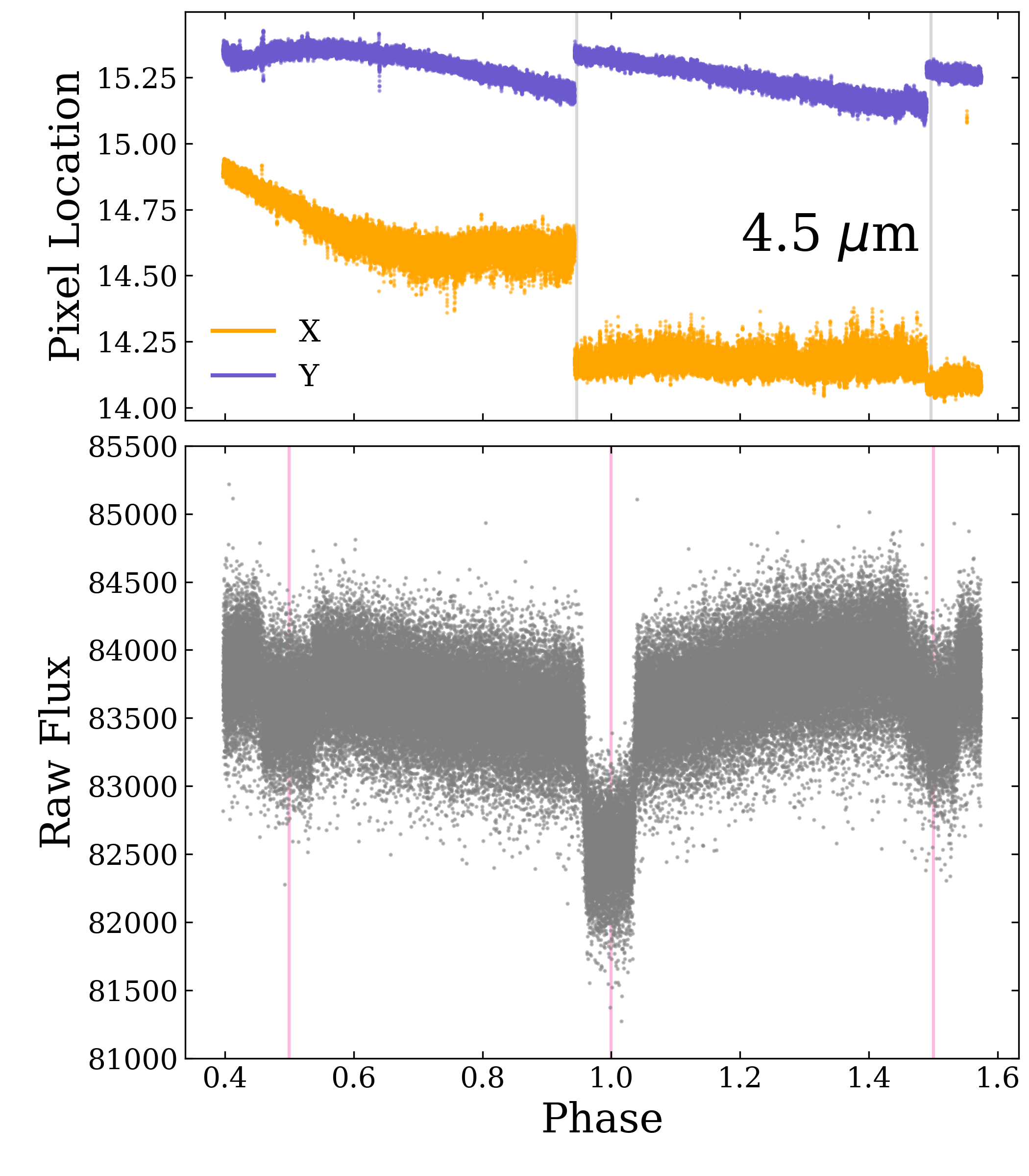}
    \caption{Centroids and raw flux from all WASP-76b phase curves. In all columns, the top panel shows the x and y centroids throughout the observation (vertical grey lines denote the AOR gaps), while the bottom panel shows the corresponding raw flux (vertical pink lines denote the expecte times of eclipses and transit). This demonstrates the strength of the intrapixel effect at 3.6 $\micron$ compared to 4.5 $\micron$, and the particular difficulties of the first 3.6 $\micron$ visit due to jumps in centroids between AORs affecting the raw flux count.}
    \label{fig:centroids}
\end{figure*}

Figure \ref{fig:centroids} shows the changes in x and y centroids over the course of all three phase curve observations, as well as the corresponding raw flux. The first 3.6 $\micron$ phase curve is split into three astronomical observation requests (AORs) with the gaps occurring just before the start of transit and in the middle of the second eclipse, showing significant jumps in centroids at these gaps. Because the eclipses are crucial anchoring points, and large gaps in centroids can be difficult to detrend due to strong intrapixel effects, particularly with the 3.6 $\micron$ channel, a second 3.6 $\micron$ phase curve was observed to ensure no AOR gap occurred near the eclipse events. This second phase curve is similarly split into three AORs, however there are no significant jumps in the centroid position at the AOR gap. The 4.5 $\micron$ phase curve is also split into three AORs with a significant gap between the first two, resulting in two groups of non-overlapping AORs. However, our detrending method discussed below is well suited to address this.

\begin{deluxetable}{c c c c}
    \tablecolumns{6}
    \tabletypesize{\footnotesize}
    \tablecaption{Observational Details}
    \label{table:observations}
    \tablehead{
    \colhead{Label} &
        \colhead{wa076bo11} &
        \colhead{wa076bo12} &
        \colhead{wa076bo21} }
    \startdata
        \hline
        Obs. Date & May 03-05 `17    &   Apr 21-23 `18  &   Apr 15-17 `17  \\
        Duration (hrs)  &   51.26   &   52.16   &   51.04   \\
        Frame Time (s)  &   0.4 &   0.4 &   2.0 \\
        Total Frames    &   430,912 &  430,912  &   90,944  \\
        Channel ($\mu$m)    &   3.6 &   3.6 &   4.5 \\ \hline
    \enddata
    \tablecomments{Label denotes the planet (wa076b), type of observation (o=orbit), Spitzer IRAC channel (1 or 2) and visit number (1 or 2)}
\end{deluxetable}

\subsection{Data Analysis}
\subsubsection{Initial Data Reduction}
We use the Photometry for Orbits, Eclipses, and Transits \citep[POET,][]{Campo2011,Stevenson2012, Cubillos2013} pipeline for data reduction and analysis, including recent updates from \cite{May2020c} to improve systematic modeling at 4.5 $\micron$. 

Centroiding of each frame is done using a 2D Gaussian fit following the suggestions of \cite{Lust2014}. Both wavelength channels are extracted using a fixed aperture size optimized for signal difference to noise ratio (SDNR). For this, we test apertures between 2.0 and 4.0 pixels in 0.25 pixel increments. Background subtraction is done in a fixed annulus between 7 and 15 pixels away from the centroids. We select an aperture of 2.25 pixels for the 3.6 $\micron$ channel and 3.0 pixels for the 4.5 $\micron$ channel.

\subsubsection{The Intrapixel Effect}
\begin{figure*}
    \centering
    \includegraphics[width = 0.335\textwidth]{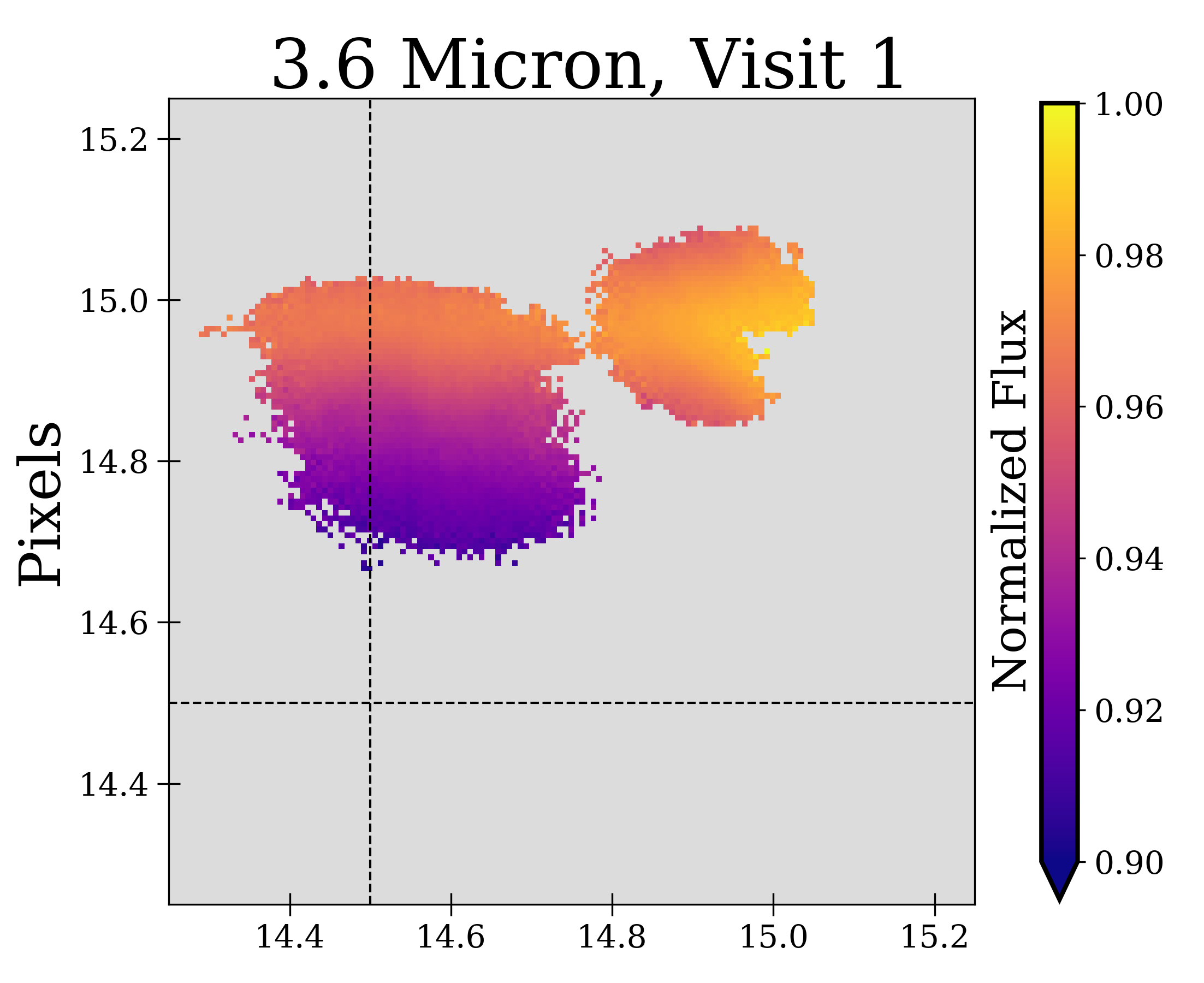}
    \includegraphics[width = 0.320\textwidth]{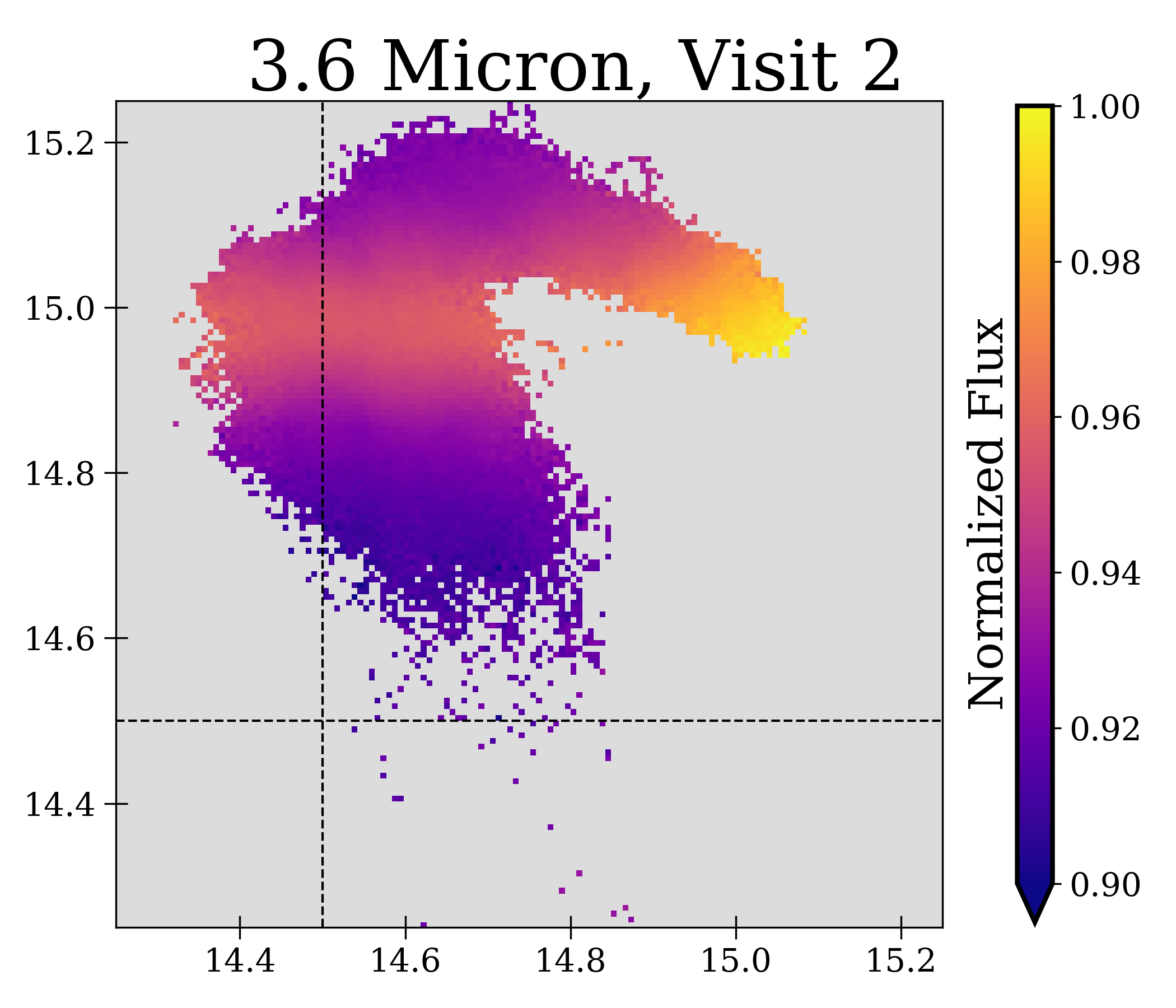}
    \includegraphics[width = 0.320\textwidth]{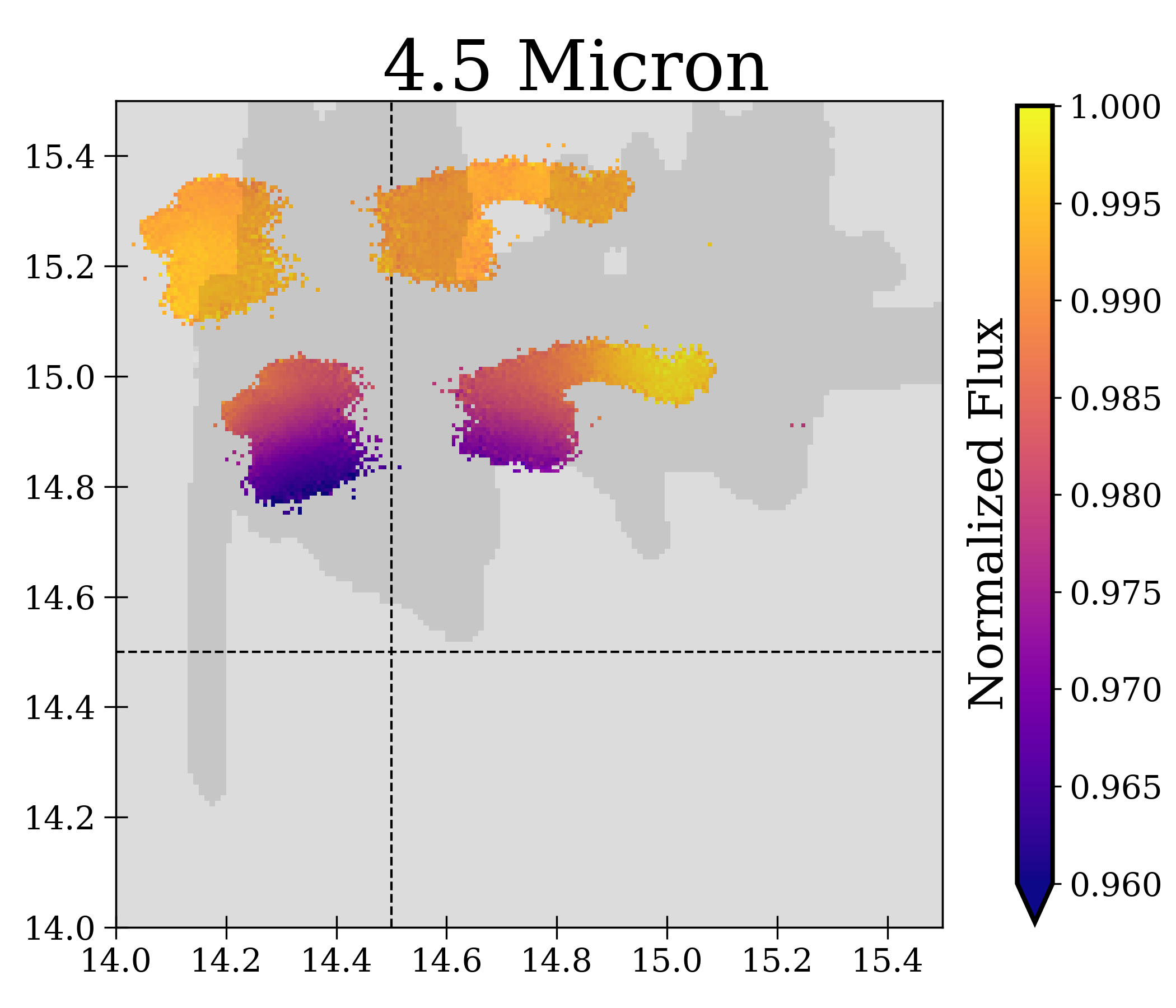}
    \caption{BLISS maps from our 3 WASP-76b phase curves. For the 3.6 $\micron$ panels these are the results of our joint fit; however, the BLISS maps were generated independently to account for temporal variability of the intrapixel sensitivity in this channel. For the 4.5 $\micron$ panel, we include the primary contribution map (lower centroid clusters) and the companion contribution map (upper centroid clusters). Because the companion is fainter than the primary, the variations around the mean of the companion map are significantly smaller. This panel also shows the extent of our fixed sensitivity map at 4.5 $\micron$ in the shaded region. In all panels the dashed lines denote the edges of a pixel. Axis labels are all in sub-pixel units.}
    \label{fig:BLISSmaps}
\end{figure*}

Systematics are modeled using Bilinearly Interpolated Subpixel Sensitivity (BLISS) mapping \citep{Stevenson2012} to address the dominant source of Spitzer IRAC systematics at 3.6 and 4.5 $\micron$. As explored in \cite{May2020c}, BLISS mapping is degenerate with point response function at full width half max (PRF FWHM) detrending, which accounts for variations in the shape of the PRF as the centroid moves towards the edges of the pixel. This effect is exacerbated by temporally binning the data, so we choose to not perform any binning in this analysis. For the 4.5 $\micron$ phase curve, we use the fixed intrapixel sensitivity map presented by \cite{May2020c} with a modification described below to address the nearby companion of WASP-76. Because such a map does not yet exist at 3.6 $\micron$, we instead perform a joint fit of both 3.6 $\micron$ phase curves using standard BLISS mapping techniques. Figure \ref{fig:BLISSmaps} shows our best fit BLISS maps for both 3.6 $\micron$ visits, as well as the dual-map results at 4.5 $\micron$ as further addressed below. 

We note that the fixed sensitivity map method used at 4.5 $\micron$ provides particularly strong constraints on phase curve parameters (most notably, the phase offset) when a significant portion of the centroids overlap with the fixed map, as is the case for WASP-76b. Similarly, at 3.6 $\micron$, because we perform a joint fit of both 3.6 $\micron$ visits, there are smaller uncertainties than a single visit fit on the phase curve parameters (again, most notably for the phase offset) due to this additional constraints of the multiple visits. 

\subsubsection{Astrophysical Source Models}
For all phase curves we model the planet's emission with spherical harmonics using {\tt{SPIDERMAN}} \citep{Louden2018}, with both 2$^{nd}$ order (symmetric) and 3$^{rd}$ order (asymmetric) models considered. For the 2${nd}$ order case, we only consider the two terms that create a dipole centered at a latitude and longitude of zero degrees, and use \texttt{SPIDERMAN}'s longitude offset parameter. The transit events are modeled using {\tt{BATMAN}} \citep{Kreidberg15} and eclipses are modeled following the formalism of \cite{MandelAgol2002ApJtransits}. We apply quadratic limb darkening for the transit events calculated using {\tt{ExoCTK}}'s limb darkening tool and Kurucz stellar models \citep{Kurucz2004} with stellar values of T$\mathrm{_{eff}}$ = 6250 K, logg = 4.13, and [Fe/H] = 0.23 \citep{West:2016aa}. At 3.6 $\micron$, the two limb darkening parameters used are [0.08, 0.13] and  at 4.5 $\micron$ they are [0.081, 0.097]. In addition to the astrophysical signals and aforementioned BLISS mapping, we consider the addition of a temporal ramp applied across the entire data set. At 4.5 $\micron$ we compare a linear temporal ramp to a no-ramp model (here a quadratic ramp produces unphysical results, namely negative nightside flux values.). At 3.6 $\micron$ we consider combinations of ramps in our joint fit: both visits with no ramp, both visits with linear ramps, both visits with quadratic ramps, and a mixed-ramp case where the first visit has a quadratic ramp and the second visit has a linear ramp. Our choice of the mixed ramp case in our joint 3.6 $\micron$ fit reflects the unreliability of the first visit due to the start and end of the visit experiencing strong changes in the measured flux (See visit 1 panels in Figure \ref{fig:centroids} and Figure \ref{fig:ch1_bestfit}), which affects the results without removing these points. Notably, we find the exact amount trimmed from the start and end of the first 3.6 $\micron$ phase curve heavily affects the shape of the resulting phase curve model when using a linear ramp, but that this can be constrained by performing the joint fit. Specifically, we tested trimming 30, 60, and 90 minutes of data from the start and end of the first 3.6 $\micron$ visit and found that the resulting nightside flux heavily depended on the amount trimmed. For that reason we decided to only consider the joint fit in this work. Our mixed temporal ramp (quadratic for visit 1 and linear for visit 2) test was to add additional flexibility in the fit of the first visit while allowing the second visit to more tightly constrain the possible phase curve. Our final fit does not trim any data from the start or end of the first 3.6 $\micron$ phase curve. 

Our best fit combination of the above models is selected using the Bayesian Information Criterion \citep[BIC,][]{Liddle2007}. These best fit models are computed using a Levenberg-Marquardt minimizer with parameter uncertainties estimated using a custom Differential-Evolution Markov Chain algorithm \citep{terBraak2008}. Simultaneously, parameters that describe the BLISS maps (subpixel grid size and minimum number of data points in a map grid) are chosen by comparing model fits to a nearest neighbor approach and selecting the step size that produces the best BIC and SDNR, respectively. 

Table \ref{table:bestfits} overviews the results of our various model combinations for both channels. The combination with the best BIC for each observation was identified as our `best-fit' model and all subsequent discussions use that model combination. While not shown in the table for simplicity, we also consider the addition of PRF FWHM detrending. As also found in \cite{May2020c}, the use of our fixed sensitivity map at 4.5 $\micron$ accounts for the PRF FWHM variations and its addition is not a preferred model solution for this data set. For the 3.6 $\micron$ phase curves we find that the addition of a 2$^{nd}$ order PRF FWHM detrending function provides improvements to the fit as demonstrated by a lower BIC value. Table \ref{table:all_params} includes the best fit (or adopted) parameters for all values included in the total functional form of the fit.

\subsubsection{Nearby Companion}
WASP-76 has a nearby companion 0.4438 $\pm$ 0.0053'' away with confirmed common proper motion \citep{Wollert2015,Ginski2016,Ngo2016, Bohn2020, Southworth2020}. This separation places both WASP-76 and its companion well within a single Spitzer pixel. The system's $\Delta$K$_{\mathrm{mag}}$ of 2.30 $\pm$ 0.05 suggests that the companion is a late G or early K star and gives a flux differential within the Spitzer 4.5 $\micron$ band of $\sim$ 5.5. This relative closeness of the two components compared to the Spitzer PSF and large flux differential prevents a standard double-Gaussian aperture fit to account for the secondary component. However, because we use Gaussian centroiding and not, for example, center of light centroiding, we do not expect our WASP-76 centroids to be affected by the significantly fainter companion.  We therefore use the position angle and offset calculated in \cite{Wollert2015} and \cite{Ginski2016} to determine the location of the companion relative to our measured
WASP-76 centroids in Spitzer pixel coordinates and compare to HST images of the system that resolve the two components to ensure we have correctly identified the relative location of the companion in our Spitzer photometry in RA and DEC space. We consider both the dilution of eclipse depths and the intrapixel effect due to the companion, as addressed in the following paragraphs.

Following \cite{Stevenson2014a} and \cite{Stevenson2014b}, we calculate the wavelength dependent dilution effect for both 3.6 and 4.5 $\micron$. Previously resolved imaging of the system from \cite{Ngo2016} report the companion's surface gravity and mass (log(g) = 4.608 cm/s$^{2}$, M = 0.712 M$_{\odot}$), giving a companion radius of 0.695 R$_{\odot}$. We adopt the stellar log(g), mass, and radius of WASP-76 from \cite{West:2016aa}. The dilution factors at 3.6 and 4.5 $\micron$ are calculated to be 0.0936 $\pm$ 0.002 and 0.0901 $\pm$ 0.003, respectively. 

To address the intrapixel effect of the companion star, at 3.6 $\micron$ we perform our standard fits such that our resulting BLISS map now encapsulates the flux from both stellar components. After obtaining the best fit, we perform a contamination correction based on the above dilution factors. However, at 4.5 $\micron$ our use of a fixed sensitivity map means we must account for each stellar component independently by performing a two-map fit. This process requires a two step approach: (1) First, we remove the fixed intrapixel sensitivity map component centered at the location of WASP-76 such that we are left only with the flux variations due to the shifting location of the companion and the planet's phase curve. (2) Second, we shift the residual flux's centroids to the companion's location and rescale the variations in our fixed sensitivity map using the flux differential between the two components in the 4.5 $\micron$ band. At this point, we run our standard fitting techniques on the residual data which produces an intrapixel map at the companion's location and the planet's best fit phase curve. The centroids at the location of the companion lay approximately half on and half off of our fixed intrapixel sensitivity map, so it is partially constrained by this overlap. The rightmost panel of Figure \ref{fig:BLISSmaps} shows the maps from this dual map fit. The shaded grey region denotes where our fixed intrapixel sensitivity map data exists, which is an extension of that in \cite{May2020c}. Both components are shown on the same scale, highlighting that the companion contribution is significantly smaller than that of the primary, but still non-negligible. We then perform a contamination correction based on the above dilution factors.

\begin{deluxetable}{l l l l l }
    \tablecolumns{6}
    \tabletypesize{\footnotesize}
    \tablecaption{Light Curve Models}
    \label{table:bestfits}
    \tablehead{
        \colhead{Label} &
        \colhead{Systematic} &
        \colhead{Ramp} &
        \colhead{Phase} &
        \colhead{$\Delta$BIC} \\ 
          & \colhead{Model} & \colhead{Model}  & \colhead{Model} & 
                }
    \startdata
        wa076bo1*   & BLISS             &  ...                      & Symm.         &  11819.8   \\
                    & + 2$^{nd}$ order  & (L. + L.)             & Symm.         &  124.9      \\
                    &   PRF-FWHM        & (Q. + Q.)           & Symm.         &  9.8     \\
                    &                   & \textbf{(Q. + L.)}   & \textbf{Symm.} &  \textbf{0.0}     \\
                    &                   &  ...                      & Asymm.        &  1196.0 \\
                    &                   & (L. + L.)             & Asymm.        &  137.9      \\
                    &                   & (Q. + Q.)           & Asymm.        &  22.8      \\
                    &                   & (Q. + L.)            & Asymm.        &  13.0     \\ \hline
        wa076bo21   & Fixed Map         &  ...      & Symm. &  79.66    \\
                    & + BLISS           & \textbf{Linear}    & \textbf{Symm.} &  \textbf{0.0}  \\
                    &                   &  ...      & Asymm. &  91.06    \\
                    &                   & Linear    & Asymm. &  11.41    \\ \hline
    \enddata
    \tablecomments{For the joint ramp fits, L stands for linear and Q stands for quadratic.The bolded rows represent the best-fit model combinations, as denoted by the $\Delta$BIC of 0.0.}
\end{deluxetable}

%
\subsection{3.6 $\micron$ Phase Curves}
\begin{figure*}
    \centering
    \includegraphics[width=0.99\textwidth]{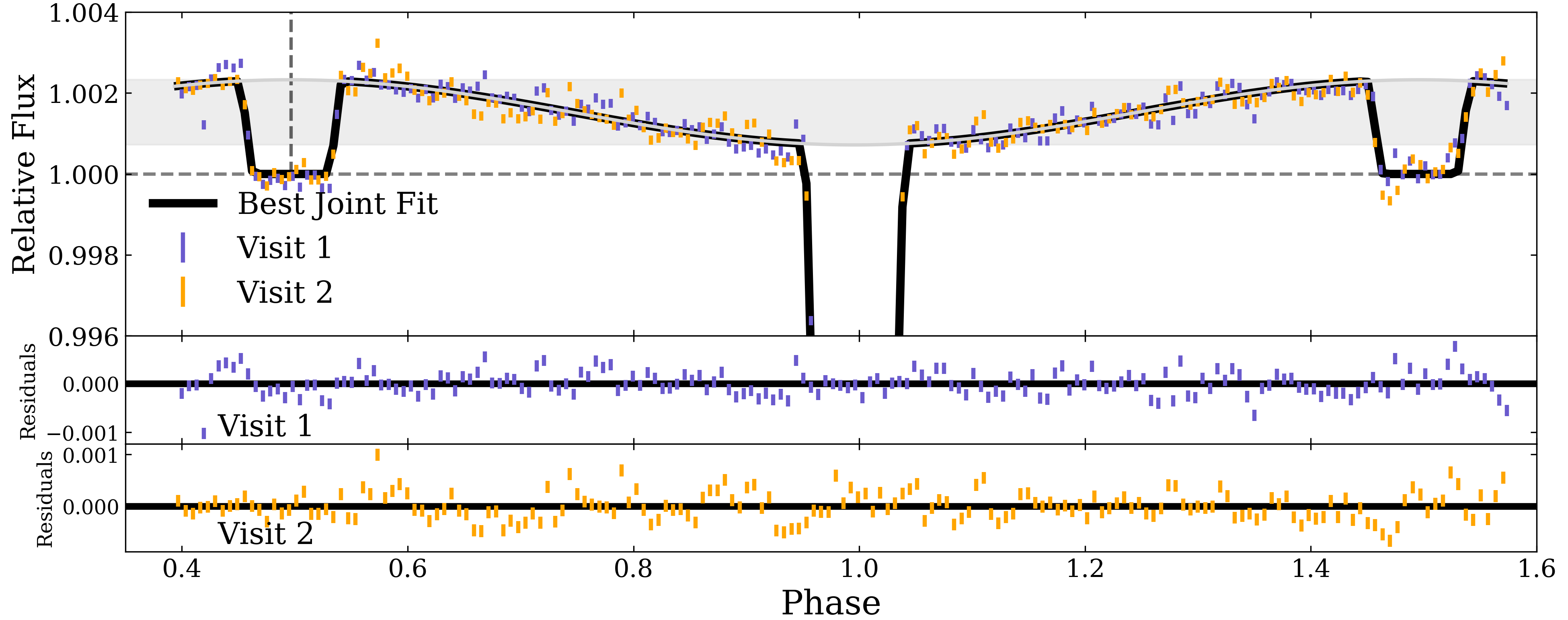}
    \caption{3.6 $\micron$ best joint fit. The horizontal shaded region denotes the maximum to minimum of the phase function to guide the eye, while the vertical dashed line denotes the peak of the phase function.}
    \label{fig:ch1_bestfit}
\end{figure*}

For parameter estimation, we perform a joint fit of both 3.6 $\micron$ phase curves, requiring that the transit depth and phase function parameters are the same for both visits. Because the first visit is split into two non-overlapping AORs and the second visit has a slight drift during one AOR, our joint fit requires both observations to have the same phase and occultation parameters and serves to constrain the independent BLISS maps for both visits. Our joint best fit model compared to the two visits is shown in Figure \ref{fig:ch1_bestfit}.

For our 3.6 $\micron$ phase curves, our joint best fit has a hot spot offset of 0.68 $\pm$ 0.48$^{\circ}$ with a phase amplitude of 804.0 $\pm$ 42.5 ppm. We measure a contamination corrected eclipse depth of 2539 $\pm$ 30 ppm, corresponding to a disk integrated day side brightness temperature of 2471 $\pm$ 27 K. The disk integrated night side brightness temperature is measured to be 1518 $\pm$ 61 K. Performing a joint fit on two 3.6 $\micron$ phase curves allows us to reach a higher precision on these parameters while tightening the constraint on the allowed BLISS maps by tying the phase curve parameters between visits. Brightness temperatures were calculated based on an interpolated Kurucz stellar model with stellar parameters from \cite{West:2016aa} as discussed above.

\subsection{4.5 $\micron$ Phase Curves}
\begin{figure*}
    \centering
    \includegraphics[width=0.99\textwidth]{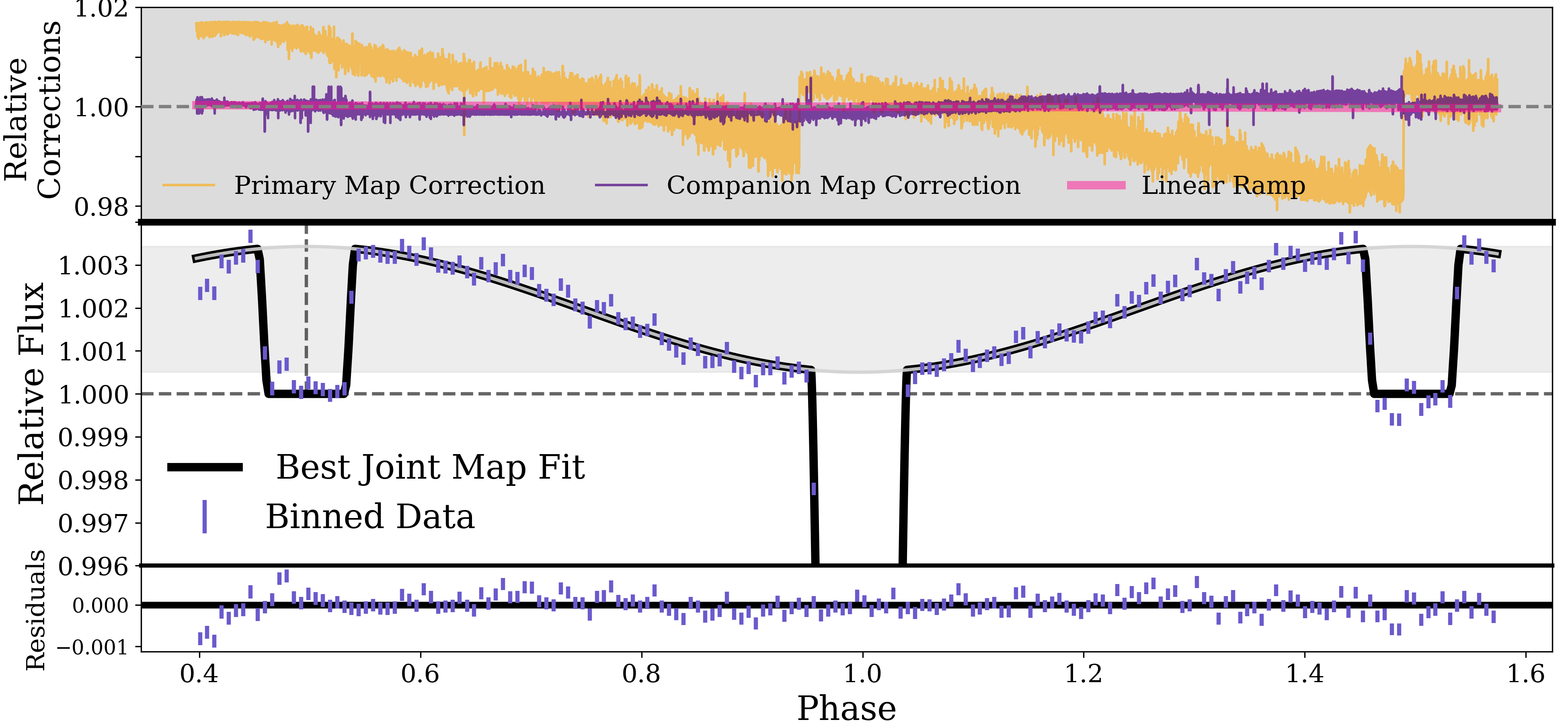}
    \caption{4.5 $\micron$ best fit. \textbf{Top:} the intrapixel map components compared to the temporal ramp. The largest effect is due to WASP-76 (orange), with a smaller effect due to the companion star (5.548 times fainter, purple). \textbf{Middle:} Data with uncertainties and best fit model. The horizontal shaded region denotes the maximum to minimum of the phase function to guide the eye, while the vertical dashed line denotes the peak of the phase function. \textbf{Bottom:} Best fit model residuals.}
    \label{fig:ch2_bestfit}
\end{figure*}

As addressed above, we perform a two-map fixed sensitivity map removal to address the intrapixel effect at 4.5 $\micron$ due to WASP-76 and its nearby companion star. 99.9\% of the centroids for WASP-76 fall on the fixed sensitivity map, while the companion star centroids fall approximately half-on and half-off of its shifted fixed sensitivity map. We allow these regions off the map to fit using a standard BLISS mapping routine. Our best fit is shown in Figure \ref{fig:ch2_bestfit}. The top panel denotes the relative contribution from the two map components. 

With WASP-76's centroids falling nearly entirely on the fixed sensitivity map, the shape of the phase curve is highly constrained by the pre-determined systematic model. This allows us to place a strong constraint on the hotspot offset of 0.67 $\pm$ 0.2$^{\circ}$ with a phase amplitude of 1464 $\pm$ 38 ppm. We measure a contamination-corrected eclipse depth of 3729 $\pm$ 52 ppm, corresponding to a day side disk integrated brightness temperature of 2699~$\pm$~32~K.  The disk integrated night side brightness temperature is measured to be 1259 $\pm$ 44 K.

\section{Numerical models of the atmospheric circulation of WASP-76\lowercase{b}}
\label{sec:GCM}
\subsection{Model setup}
To simulate the atmospheric circulation of WASP-76b, we solve the primitive equations of meteorology with the {\tt MITgcm} \citep{Adcroft:2004} coupled to double-gray radiative transfer with the {\tt TWOSTR} \citep{Kylling:1995} package of {\tt DISORT} \citep{Stamnes:2027}. This model has previously been applied to study the atmospheric circulation of hot Jupiters and brown dwarfs \citep{Komacek:2017,Komacek:2019aa,Komacek:2020aa,Steinrueck:2020aa,Tan:2020aa,Tan:2021aa}, and it has been modified to take into account the effects of hydrogen dissociation and recombination on the thermodynamics of ultra-hot Jupiter atmospheres \citep{Tan:2019aa,Mansfield:2020aa}. This is an intermediate-complexity GCM that builds on the heritage of previous work studying gaseous exoplanet atmospheres with GCMs that utilize double-gray radiative transfer (e.g., \citealp{Dobbs-Dixon:2010aa,Heng:2011a,perna_2012,Rauscher_2012,Rauscher_2013,Rauscher:2014,Rauscher:2017wm,Zhang:2017b,Mendonca:2018ab,Flowers:2018aa,Roman:2019aa,Dietrick:2020aa,May:2020vr,Mendonca:2020aa,Roman:2020aa,Beltz:2021aa,Harada:2021tc}). Generally, double-gray GCMs provide broadly similar predicted phase curve properties as more complex non-gray models \citep{Parmentier:2020aa}. In this work, we utilize such an intermediate-complexity model in order to conduct a broad suite of simulations to scan possible parameter space. As discussed in Section \ref{sec:postprocess}, we then post-process our GCM output with a state-of-the-art radiative transfer code \citep{Zhang:2017b} in order to compare with \textit{Spitzer} observations.

The GCM assumes an atmosphere that is composed purely of hydrogen. We include the effects of hydrogen dissociation and recombination by tracking the local atomic hydrogen mass mixing ratio, which is treated as a passive tracer that is dynamically coupled to the atmospheric flow as in \cite{Tan:2019aa}. We then include local cooling in regions where the atomic hydrogen fraction increases (and local heating where the atomic hydrogen fraction decreases). This atomic hydrogen tracer is forced to relax quickly toward the expected local chemical equilibrium abundance given by the Saha equation \citep{Berardo:2017aa,Bell:2018aa}. In order to both maintain numerical stability and keep the mixing ratio of atomic and molecular hydrogen close to chemical equilibrium values, as in \cite{Tan:2019aa} we set the relaxation timescale of the atomic hydrogen tracer to be 1.5 times the dynamical time step.

As discussed in the introduction, WASP-76b lies in a region of parameter space in which both large-scale drag due to magnetic effects and/or large scale turbulence and deposited heating in its interior may affect its atmospheric circulation. As a result, we conducted two coupled parameter sweeps covering a range of atmospheric frictional drag and internal heating. First, we crudely represented the effects of Lorentz forces \citep{Perna_2010_1,perna_2012,batygin_2013,Rauscher_2013,Rogers:2020,Rogers:2014,Hindle:2019aa} and/or large-scale turbulence \citep{Li:2010,Youdin_2010,Fromang:2016} as a Rayleigh drag (i.e., linear in wind speed) characterized by a spatially constant drag timescale $\tau_\mathrm{drag}$, as done previously (e.g., \citealp{Perez-Becker:2013fv,Komacek:2015,Arcangeli:2019aa,Tan:2019aa}). As in \cite{Mansfield:2020aa}, we vary the drag timescale over a wide range (from $10^3~\mathrm{s}$ to $10^7~\mathrm{s}$) to encompass the uncertainties in the magnetic field strength \citep{Yadav:2017} and scale of turbulent dissipation \citep{Koll:2017} in the atmosphere of WASP-76b. 

Second, because WASP-76b is both low-gravity and receives strong stellar irradiation, it likely undergoes intense deposited heating that acts to inflate its radius \citep{Thorngren:2017}. This deposited heating may affect the atmospheric structure of the planet by heating the deep atmosphere and forcing a shallower radiative-convective boundary \citep{Sainsbury-Martinez:2019aa,Thorngren:2019aa,Carone:2019aa,Bayens:2021uo,Sarkis:2021aa}. We conducted simulations for two end-members of internal heating.  The first case assumes a forced thermal heat flux from the bottom boundary (100 bars) of $F_\mathrm{int}  = 3543.75~\mathrm{W}~\mathrm{m}^{-2}$, corresponding to a  temperature $T_\mathrm{int} = (F_\mathrm{int}/\sigma)^{1/4}$ of $500~\mathrm{K}$, which lies at a similar order of magnitude to the value commonly assumed in GCMs of hot Jupiters that include an interior heat flux (e.g., \citealp{Rauscher_2012,Showman:2014,Komacek:2017,Mayne:2017,Mendonca:2020aa,Roman:2020aa}). Note that the strength of convective perturbations in gas giant atmospheres is expected to be relatively isotropic, with only a slight increase in heat flux at the poles \citep{Showman:2013}. As a result, we assume in all GCMs presented here that the heat flux at the bottom of the domain is horizontally isotropic.

We also considered the end-member of an extremely hot planet with a deep adiabatic interior,  with a thermal heat flux from the bottom boundary of $F_\mathrm{int} = 4.474 \times 10^{7}~\mathrm{W}~\mathrm{m}^{-2}$ corresponding to a temperature $T_\mathrm{int} = 5300~\mathrm{K}$, chosen to lead to a nearly-uniform atmospheric temperature structure at pressures above $\sim 1~\mathrm{bar}$. Note that this value is larger than in one-dimensional models that include hot interiors \citep{Gao:2020aa}, because the GCM used here includes a weak bottom drag as in \cite{Komacek:2017} that extends from 100 bars to 10 bars, placed to ensure insensitivity to initial conditions \citep{Liu:2013}. As a result, here we choose the value of interior temperature to force the pressure at which the atmosphere becomes horizontally uniform to be comparable to the radiative-convective boundary expected from one-dimensional atmospheric structure models \citep{Thorngren:2019aa,Gao:2020aa}. Future work will include long timescale model integrations and convective parameterizations in order to explore the detailed effect of the internal heat flux on the deep atmospheric circulation of hot and ultra-hot Jupiters. 

Table \ref{table:params} shows our assumed model parameters for the suite of GCMs with varying drag timescale and interior temperature presented in this work.  
\begin{table*}
\begin{center}
\begin{tabular}{ l  l  l }
\hline
{\bf Parameter} & {\bf Value} & {\bf Units} \\
\hline
Radius & 1.85 & $R_\mathrm{Jup}$ \\
Gravity & 6.46 & m~s$^{-2}$\\
Rotation period & 1.81 & Earth days \\
Specific heat capacity of H$_2$ & $1.300 \times 10^4$ & J~kg$^{-1}$~K$^{-1}$ \\
Specific heat capacity of H & $2.079 \times 10^4$ & J~kg$^{-1}$~K$^{-1}$   \\
Specific gas constant of H$_2$ & 3714 & J~kg$^{-1}$~K$^{-1}$ \\
Specific gas constant of H & 5940 & J~kg$^{-1}$~K$^{-1}$ \\
Irradiation temperature & 3150.87 & K \\
Interior temperature & $\left[500, 5300\right]$ & K \\
Visible opacity & $10^{0.0478(\mathrm{log}_{10}p[\mathrm{Pa}])^2 - 0.1366(\mathrm{log}_{10}p[\mathrm{Pa}]) - 3.2095}$ & m$^2$~kg$^{-1}$ \\
Infrared opacity & $\mathrm{max}\left(10^{0.0498(\mathrm{log}_{10}p[\mathrm{Pa}])^2 - 0.1329(\mathrm{log}_{10}p[\mathrm{Pa}]) - 2.9457}, 10^{-3}\right)$ & m$^2$~kg$^{-1}$ \\ 
Drag timescale & $\left[10^3,10^4,10^5,10^6,10^7\right]$ & s\\
Horizontal resolution & C48 & - \\
Vertical resolution & 70 & layers\\
Lower boundary & 100 & bars \\
Upper boundary & 10 & $\mu$bars \\
Dynamical time step & 10 & s \\
Radiative time step & 20 & s \\
H tracer relaxation timescale & 15 & s \\
\hline
\end{tabular}
\caption{\textbf{Planetary and atmospheric properties assumed for GCM simulations of WASP-76b.} Brackets indicate parameter sweeps performed over ranges of interior temperature and drag timescale.}
\label{table:params}
\end{center}
\end{table*}
Specifically, we perform simulations for five different drag timescales covering $10^3-10^7~\mathrm{s}$ with two different assumptions of the interior temperature (which we term the ``cold interior'' and ``hot interior'' cases), resulting in a grid of 10 GCMs. We use the pressure-dependent double-gray opacity formulation similar to  \cite{Tan:2019aa}, and additionally impose a minimum opacity in the planetary thermal band of  $10^{-3}~\mathrm{m}^2~\mathrm{kg}^{-1}$, equivalent to the constant opacity used in the previous 1D and 3D double-gray modeling work of \cite{Guillot:2010} and \cite{Rauscher_2012}. As in \cite{Tan:2021aa}, we include this minimum opacity to ensure that the upper levels of the model at pressures $\lesssim 1.5~\mathrm{mbar}$ where this opacity is larger than our pressure-dependent opacity are radiatively active.  The horizontal resolution of the GCM is C48, which is  equivalent to $192 \times 96$ grid points in longitude and latitude. The GCM has 70 vertical layers, equally spaced in log-pressure from 100 bars to 10 $\mu$bars. Each GCM simulation is initialized from a rest state with an isothermal atmosphere at $T = 2200~\mathrm{K}$. Each simulation is then run until it reaches an equilibrated state in domain-integrated (i.e., extending from the top of the domain to $100$ bars) kinetic energy, which for simulations with weak frictional drag typically takes $\approx 5\rm{,}000$ Earth days of model time. All results shown are averages over the last 500 Earth days of model time.
\subsection{GCM results}
The interior temperature of hot Jupiters is linked to their atmospheric dynamics through its impact on the three-dimensional atmospheric temperature structure. Hot Jupiters that undergo deposited heating in the interior are expected to have larger internal heat fluxes, which causes the upper boundary of the convective zone (the radiative-convective boundary) to move to lower pressures \citep{Guillot_2002,Arras:2006kl}. Notably, the pressure at which the radiative-convective boundary of hot Jupiters lies is expected to vary with both longitude and latitude due to incident stellar radiation suppressing cooling from the interior on the dayside \citep{Guillot_2002,Spiegel:2013,rauscher_showman_2013}, leading to a larger interior cooling rate and shallower radiative-convective boundaries on the non-irradiated nightside than on the irradiated dayside. Recently, \cite{Thorngren:2019aa} showed that the interior heat fluxes of hot Jupiters can be large enough to cause the radiative-convective boundaries in the irradiated regions of low-gravity hot Jupiters to lie at pressures $\lesssim 1~\mathrm{bar}$. Given that these pressures approach the levels probed by low-resolution spectroscopy \citep{Dobbs-Dixon:2017aa}, the interior heat flux of hot Jupiters can potentially impact observable properties of the planetary atmosphere. 

\begin{figure*}
    \centering
    \includegraphics[height=.65\textheight]{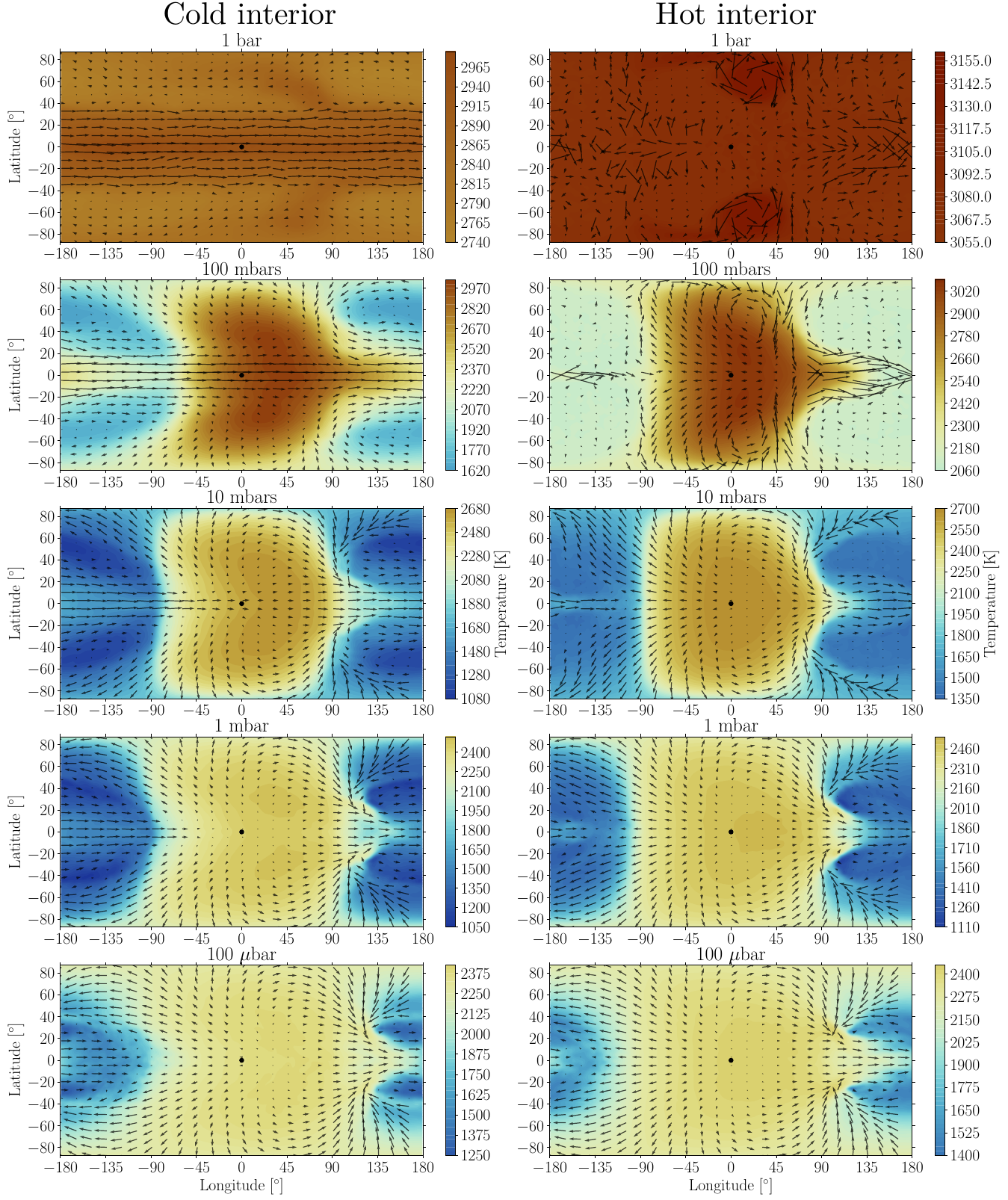}
    \caption{{\bf GCM predictions for the pressure-dependent temperature and wind pattern of WASP-76b.} Shown are temperature and wind maps with varying pressure from 1 bar to 100 $\mu$bar from our cold interior case (left column) and hot interior case (right column), both with an assumed drag timescale of $\tau_\mathrm{drag} = 10^7~\mathrm{s}$. Each map shares the same temperature color scale, and colorbars are shown for each map to display the temperature variation for each case. The normalization of the wind vectors is different for each map. The dot at the center of each map corresponds to the substellar point. We find that day-night contrasts are muted in the hot interior case, leading to weaker east-west equatorial winds and a reduced offset of the temperature maximum from the substellar point relative to the cold interior case. 
    }
    \label{fig:tempwind}
\end{figure*}

\begin{figure}
    \centering
    \includegraphics[width=.45\textwidth]{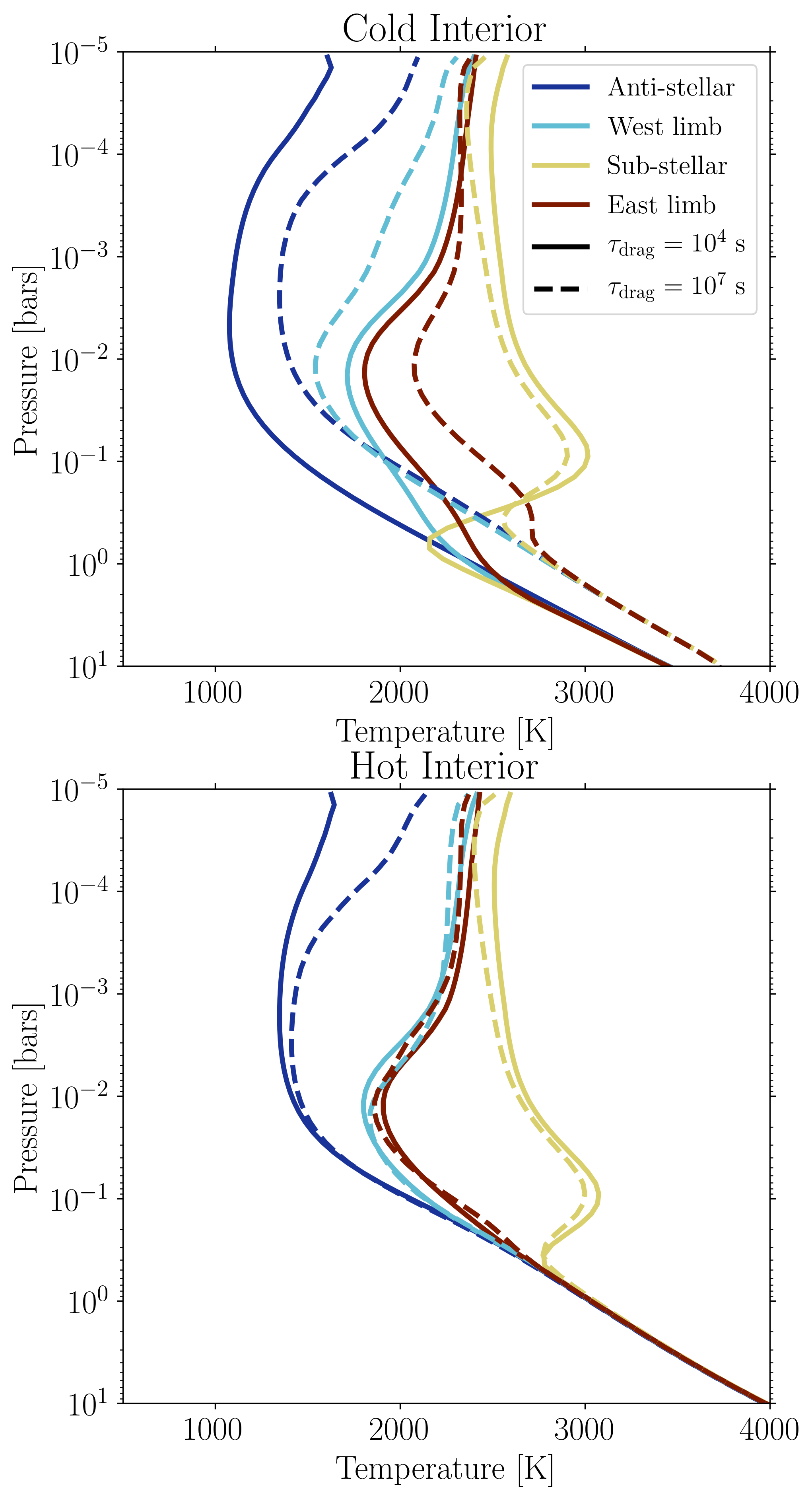}
    \caption{{\bf GCM predictions for the temperature-pressure profile of WASP-76b.} Shown are latitudinally averaged temperature-pressure profiles centered at the anti-stellar point, west limb, sub-stellar point, and east limb. Results are shown for cases with a cold interior (top panel) and hot interior (bottom panel) and for cases with strong drag characterized by $\tau_\mathrm{drag} = 10^4~\mathrm{s}$ (solid lines) and weak drag with $\tau_\mathrm{drag} = 10^7~\mathrm{s}$ (dashed lines). The contrast between the day and night side thermal profiles is largest for cases with strong drag and a cold interior. All models show a deep dayside thermal inversion, with no corresponding inversion on the nightside at similar pressures. 
    }
    \label{fig:tpgcm}
\end{figure}

The key impact of a hot internal temperature is to reduce the pressure at which the atmosphere departs from a nearly spatially uniform adiabat. As a result, a hotter internal temperature will force the temperature distribution of the deep atmosphere to be more horizontally uniform. Figure \ref{fig:tempwind} shows temperature and wind maps at various pressures from two GCM simulations of WASP-76b, both with a weak frictional drag timescale of $10^7~\mathrm{s}$ but for separate cases of a cold and hot interior temperature. Additionally, Figure \ref{fig:tpgcm} shows temperature-pressure profiles from these cold and hot interior cases with $\tau_\mathrm{drag} = 10^7~\mathrm{s}$ along with those for cases with $\tau_\mathrm{drag} = 10^4~\mathrm{s}$.  The main difference between the cases with a cold and hot interior is that the day-night temperature contrast is significantly larger in the case of a cold interior. This difference is largest at pressures of $\approx 10-100~\mathrm{mbars}$, which are near the pressure levels probed by warm Spitzer photometry. This is because in the hot interior case, the deep atmosphere at pressures $\gtrsim 1~\mathrm{bar}$ is nearly adiabatic and is characterized by small horizontal temperature contrasts. Additionally, the temperature of the deep atmosphere in the hot interior case is not affected by the assumed $\tau_\mathrm{drag}$, while in the cold interior case the deep atmosphere is warmer for longer $\tau_\mathrm{drag}$. This may be due to enhanced downward heat transport by the stronger atmospheric circulation at longer $\tau_\mathrm{drag}$ \citep{Tremblin:2017,Sainsbury-Martinez:2019aa,Carone:2019aa}, and requires further detailed study. Overall, because the pressure levels near $\sim 100$ mbar probed by Spitzer lie only a few scale heights above the deep flow, the isobaric day-to-night temperature contrasts in the observable atmosphere can be significantly reduced by a hot interior.

\begin{figure*}
    \centering
    \includegraphics[height=.65\textheight]{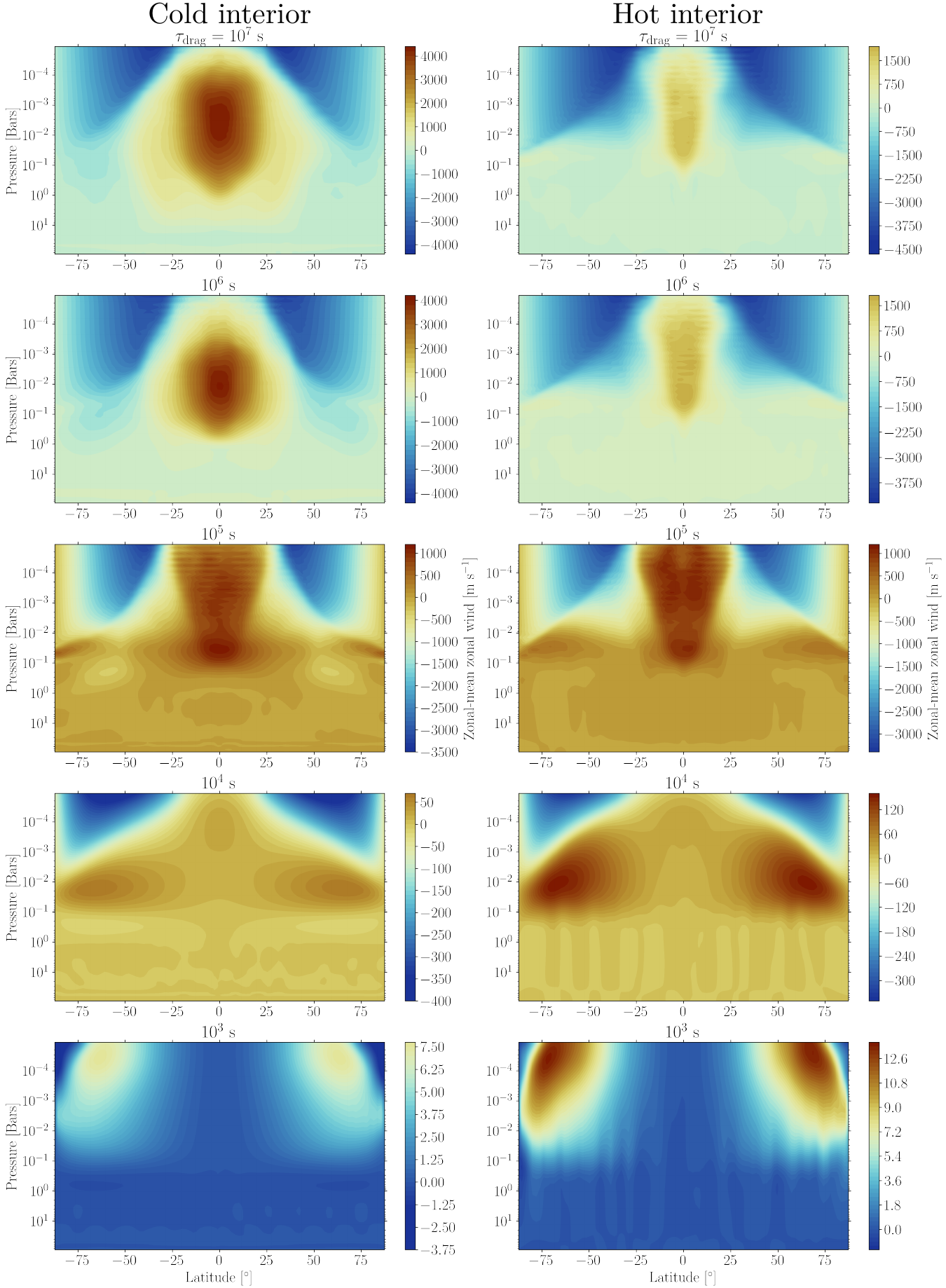}
    \caption{{\bf Longitudinal average of the eastward winds from GCM simulations with varying interior temperature and drag strength.} Shown are maps of the zonal-mean zonal wind with varying drag  timescale from $\tau_\mathrm{drag} = 10^7~\mathrm{s}-10^3~\mathrm{s}$ for the cold interior case (left column) and the hot interior case (right column). To facilitate comparison between the hot and cold interior cases, maps for a given $\tau_\mathrm{drag}$ (i.e., each row) share a wind color scale. The peak wind speed of the equatorial jet is significantly faster in simulations with a cold interior than in those with a hot interior. This is because the day-night forcing that drives equatorial superrotation is larger in the cold interior case. Strong drag prevents the formation of east-west jets in both cases, and instead the flow in cases with $\tau_\mathrm{drag} < 10^5~\mathrm{s}$ is characterized by day-to-night winds.
    }
    \label{fig:uzonal}
\end{figure*}

Critically, the decrease in the day-night temperature contrast with a hot interior reduces the amplitude of large-scale waves that are triggered by this temperature contrast \citep{Showman_Polvani_2011,Perez-Becker:2013fv}. This in turn will reduce eastward equatorial eddy momentum transport \citep{Showman_Polvani_2011,Tsai:2014,Hammond:2018aa}, leading to a decrease in the strength of equatorial superrotation at observable levels \citep{Kempton:2012aa,showman_2013_doppler} and greater influence of the divergent component of the circulation (i.e., day-night flow) \citep{Hammond:2021aa}. Figure \ref{fig:uzonal} shows the zonal (east-west) average of the zonal wind from our full suite of GCMs. As expected, in the cases with weak drag (characterized by drag timescales $\gtrsim 10^5~\mathrm{s}$) the maximum wind speed of the equatorial jet is over two times slower in the hot interior case relative to the cold interior case. Additionally, the jet is shallower in the hot interior case because the day-night forcing decays more quickly with depth than in the cold interior case.

In both cases with a cold and hot interior, the strength of applied frictional drag significantly impacts the speed of the equatorial jet. This has been found by a wide range of previous work (e.g., \citealp{Perna_2010_1,Rauscher_2013,showman_2013_doppler,Koll:2017}), and is because applied drag acts to remove momentum from the flow and convert it to heat. Notably, there is a transition from flow characterized by a strong equatorial jet with weak drag to flow characterized by day-to-night circulation at strong drag. This transition occurs in our GCMs at a drag timescale of $\sim 10^5~\mathrm{s}$, similar to that found by \cite{Perez-Becker:2013fv,showman_2013_doppler,Komacek:2015}. This is because for drag timescales $< 10^5~\mathrm{s}$, frictional drag becomes the dominant term in the momentum equation balancing the pressure gradient driven by day-night temperature contrasts \citep{Komacek:2015,Koll:2017}. As a result, though the local wind speeds in our GCMs with short frictional drag timescales can be as large as $\approx 3~\mathrm{km}~\mathrm{s}^{-1}$ for the case with $\tau_\mathrm{drag} = 10^4~\mathrm{s}$, the flow is almost uniformly eastward east of the substellar point and westward west of the substellar point. This flow geometry leads to a circulation that is not characterized by jets, but instead by substellar-to-antistellar flow. 

\begin{figure*}
    \centering
    \includegraphics[height=.65\textheight]{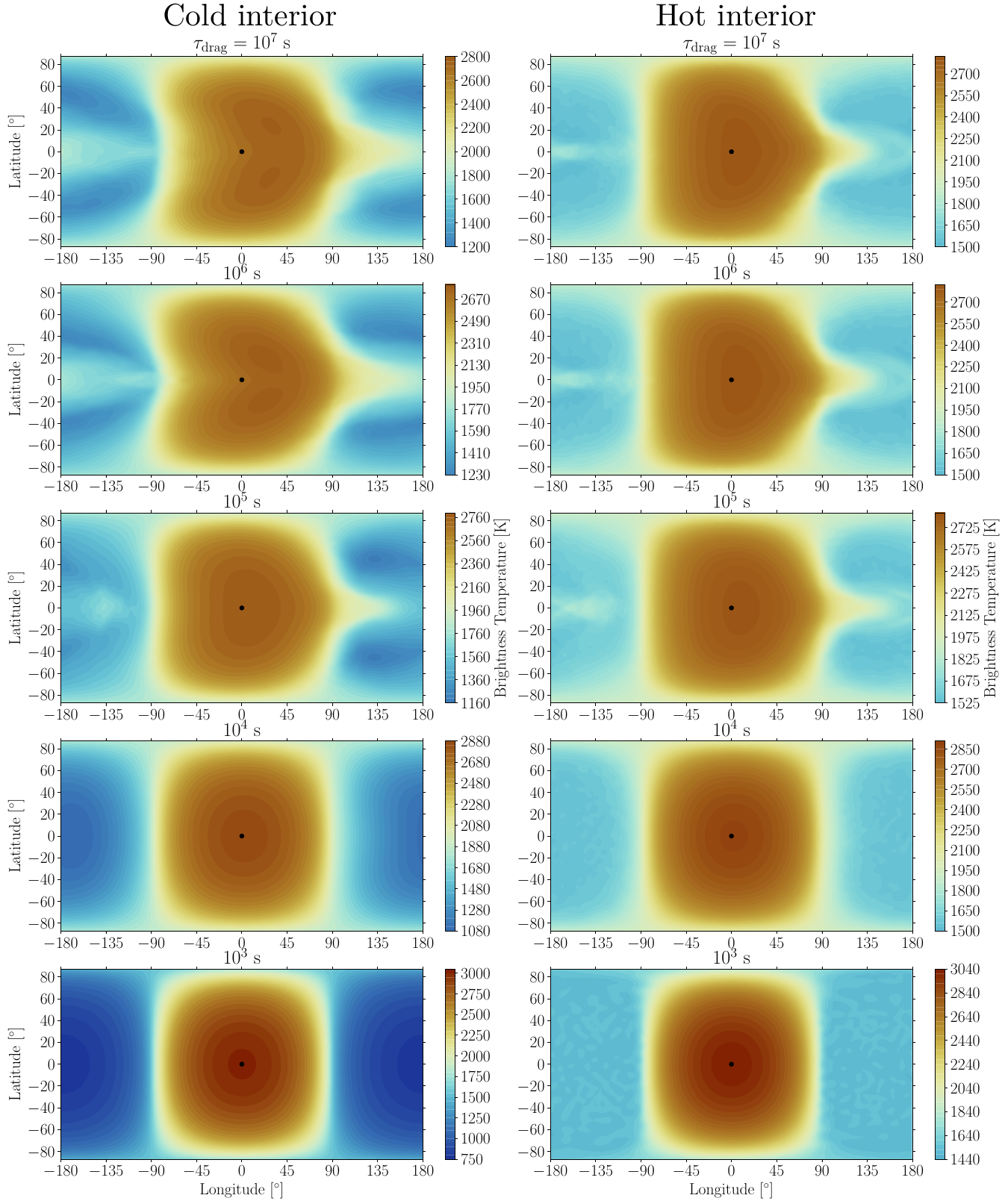}
    \caption{{\bf Maps of brightness temperature from GCM simulations with varying interior temperature and drag strength.} Shown are maps of brightness temperature with varying drag timescale from $\tau_\mathrm{drag} = 10^7~\mathrm{s}-10^3~\mathrm{s}$ for the cold interior case (left column) and the hot interior case (right column). The brightness temperature shown here is calculated from the bolometric infrared flux output from our GCM simulations. All maps share a temperature color scale in order to facilitate comparison between them, while individual colorbars display the temperature variation in each map. The brightness temperature in the standard cold interior case with weak drag has the expected chevron pattern, with a resulting eastward offset of the brightest point from the substellar point. Meanwhile, the hot interior case has a significantly reduced offset, along with a smaller contrast in brightness between the day side and night side. In both cases, the hot spot offset decreases and day-night brightness contrast increases with decreasing drag timescale. 
    }
    \label{fig:brighttemp}
\end{figure*}

The impact of interior temperature on the atmospheric temperature pattern and wind speeds has consequences for the interpretation of phase curve observations of hot Jupiters. Specifically, the interior temperature may impact both the phase curve amplitude and phase curve offset of WASP-76b. As discussed above, a hotter interior will reduce day-night temperature contrasts, leading to a reduction in the day-to-night flux contrast and resulting in a diminished phase curve amplitude. Because the reduced day-night temperature contrast leads to a slowed equatorial jet, a hotter interior will also cause the phase curve offset to decrease \citep{Cowan:2011,Showman_Polvani_2011}. Both of these effects of the interior temperature on the emitted flux distribution are evident in Figure \ref{fig:brighttemp}, which shows brightness temperature maps calculated from the bolometric infrared flux output from our GCM simulations of WASP-76b. For all drag timescales considered, the hotter interior case has a smaller brightness temperature (and emitted flux) contrast between the day side and night side hemispheres than the cold interior case. Additionally, the hotter interior cases have a smaller offset of the hottest hemisphere from the substellar point (i.e., a smaller ``hot spot'' offset) than the cold interior cases. 

The assumed atmospheric frictional drag timescale also greatly impacts the emitted flux distribution predicted by our GCMs. This is because drag acts to reduce atmospheric heat transport from day side to night side by damping the large-scale waves that act to diminish temperature variations \citep{Perez-Becker:2013fv}. As a result, increasing the strength of drag (reducing the frictional drag timescale) leads to an increased day-to-night temperature contrast and causes a resulting increase in the phase curve amplitude \citep{Komacek:2017}. Because these waves driven by the day-night temperature contrast also act to drive equatorial superrotation, increasing the strength of drag leads to a reduction in the speed of the eastward equatorial jet \citep{Rauscher:2012,Rauscher_2013,showman_2013_doppler}. As a result, increasing the drag strength leads to a reduction in the phase curve offset \citep{Komacek:2017}. Both of these effects of drag on the phase curve amplitude and offset can be seen in Figure \ref{fig:brighttemp}. In both the cold and hot interior cases, simulations with stronger drag have a a larger day-to-night brightness temperature contrast and a smaller offset of the hottest hemisphere from the substellar point. In cases with weak drag characterized by a drag timescale $\gtrsim 10^5~\mathrm{s}$, the equatorial wave pattern and resulting eastward jet cause the day side temperature structure to show a ``chevron'' pattern \citep{Showman_Polvani_2011} that is associated with a significant phase curve offset. Meanwhile, with strong drag characterized by a drag timescale $\lesssim 10^4~\mathrm{s}$, the equatorial jet does not form and as a result the brightness temperature maps shown in Figure \ref{fig:brighttemp} become nearly symmetric around the substellar point.
\section{Interpreting the Spitzer phase curve of WASP-76\lowercase{b}}
\label{sec:disc}

\subsection{Comparing our Eclipse Depths to the Literature}
Table \ref{table:eclipses} overviews the eclipse depths taken from our phase curves compared to literature values from \cite{Fu:2020aa} and \cite{Garhart20}. Our 3.6 $\micron$ depth combines 4 eclipses from 2 phase curves and our 4.5 $\micron$ depth combines 2 eclipses from 1 phase curve. Our 4.5 $\micron$ values are in agreement with previous work; however there is a significant discrepancy in our 3.6 $\micron$ values. Both \cite{Fu:2020aa} and \cite{Garhart20} use Pixel-Level Decorrelation (PLD) to address the intrapixel effect, while we only consider BLISS mapping in this work. While work by \cite{Kilpatrick2017} shows a general agreement between the Pixel Gain Map (PMAP), Nearest Neighbors, and PLD systematic removal methods for eclipses at both 3.6 $\micron$ and 4.5 $\micron$, upcoming work by Fu et al., 2021 (in preparation) finds that discrepancies can exist between PLD and BLISS mapping for eclipse and transit observations, and that eclipse and transit depths from single event observations can depend heavily on the amount of data trimmed from the start due to an initial instrument ramp. This is a significant effect at 3.6 $\micron$, regardless of the temporal ramp used to detrend the data, and is seen to go away with the use of a fixed intrapixel sensitivity map at 4.5 $\micron$. For a direct comparison, we reduced the stand-alone eclipses considered in \cite{Garhart20} using BLISS mapping instead of their PLD method. While we see the same trends as discussed above (i.e., the upcoming work from Fu et al. showing eclipse depths depend on the amount of data trimmed from the start), we can reproduce the eclipse depths of \cite{Garhart20} within 1-$\sigma$ using their same ramps and temporal trimming --- suggesting that the discrepancy between our 3.6 $\micron$ depth from the phase curves and the stand-alone eclipse in the above works is due to intrinsic difference in the eclipse depths. This is likely a result of the lower reliability of the 3.6 $\micron$ channel due to its enhanced intrapixel effect causing additional unknown systematics. Further, the additional eclipses added in \cite{Fu:2020aa} use the first eclipse of both phase curves presented in this paper, which we find are more affected by initial instrumental ramps than the second eclipse in each phase curve. 

\begin{deluxetable}{c c c c}
    \tablecolumns{4}
    \tabletypesize{\footnotesize}
    \tablecaption{Dilution Corrected Eclipse Depths}
    \label{table:eclipses}
    \tablehead{
        \colhead{ } &
        \colhead{This Work} &
        \colhead{\cite{Fu:2020aa}} &
        \colhead{\cite{Garhart20}}
        }
    \startdata
        3.6 $\micron$   &   2539 $\pm$ 30 ppm    &   2988 $\pm$ 65 ppm $^{*}$ &   2979 $\pm$ 72 ppm \\
        4.5 $\micron$   &   3729 $\pm$ 52 ppm  &   --  &   3762 $\pm$ 92 ppm \\ \hline
    \enddata
    \tablecomments{$^{*}$weighted average of the values reported. These three 3.6 $\micron$ depths reported in \cite{Fu:2020aa} include the first eclipses of both of our 3.6 $\micron$ phase curves and the stand alone eclipse analyzed in \cite{Garhart20}. All three values in these previous works are more than 1-$\sigma$ different from one another, suggesting the 3.6 $\micron$ data is unreliable. }
\end{deluxetable}

\subsection{Post-Processed Spitzer Phase Curve Models}
\label{sec:postprocess}

To compare the GCM results to the Spitzer phase curves, we ``post-process'' each of the GCM outputs with a radiative transfer (RT) code that determines the hemisphere-averaged, wavelength-by-wavelength thermal emission.  This allows us to generate thermal emission \textit{spectra} for each orbital phase.  We then integrate these spectra over the Spitzer 3.6 and 4.5 $\mu$m bandpasses to produce simulated phase curves that can be compared directly against the observational data (Figure~\ref{fig:model_data_comp}).  

The RT post-processing code that we use is described in \citet{Zhang:2017b}.  Briefly, the code solves the radiative transfer equation for the approximation of pure thermal emission along $n \times m$ individual 1-D observer-directed sightlines through the atmosphere, where $n$ is the number of latitude points and $m$ is half the number of longitude points in the GCM output (in this case 94 and 96, respectively).  The flux coming out along each ray is then multiplied by the projected area of its respective grid cell and added together to calculate the total power emanating from the visible hemisphere.  We assume an atmosphere comprised of solar composition gas in thermochemical equilibrium. 

For the current work, we have updated the opacity and chemical equilibrium data used in our RT post processing code to extend to the higher temperatures of ultrahot Jupiters like WASP-76b. Our opacity and chemical composition tables now cover the range \mbox{500 -- 5,000 K} and have been updated to include opacities and abundances from key high-temperature absorbers --- of note for this work is the inclusion of H$^{-}$, which can significantly impact the thermal emission spectra in the \textit{Spitzer} bandpasses. These updated opacities are taken from \citealt{Gordon2017} (C$_2$H$_2$); \citealt{Yurchenko2014} (CH$_4$); \citealt{Li2015} (CO); \citealt{Rothman2010} (CO$_2$); \citealt{Barber2006} (H$_2$O); \citealt{Azzam2016} (H$_2$S); \citealt{Harris2006} (HCN); \citealt{Yurchenko2011} (NH$_3$); \citealt{SousaSilva2015} (PH$_3$); \citealt{Ryabchikova2015} (TiO); \citealt{McKemmish2016} (VO); \citealt{Malik2019, Burrows2000,Burrows2003, Allard2012, Allard2016} (Na, K); \citealt{John1988} (H$^-$), and extrapolated from \citealt{Zhang2019, Zhang2020} for H and H$_2$. HELIOS-K \citep{Grimm:2021aa} is used to generate molecular opacities from the molecular line lists, and GGChem \citep{Woitke2018} is used to calculate the resulting chemical equilibrium data. We also now account for the thermal dissociation of H$_2$ when calculating the vertical altitude ($z$) coordinate of the atmosphere based on hydrostatic equilibrium, using the H/H$_2$ fractions output by the GCM.  This involves calculating the mean molecular weight separately and self-consistently in each grid cell of the GCM, prior to performing the hydrostatic equilibrium calculation for the $z$-coordinates.  This step is critical for correctly implementing the line-of-sight ray striking geometry in our RT calculations.

\subsection{Model comparisons}
\begin{figure*}
    \centering
    \includegraphics[width=0.49\textwidth]{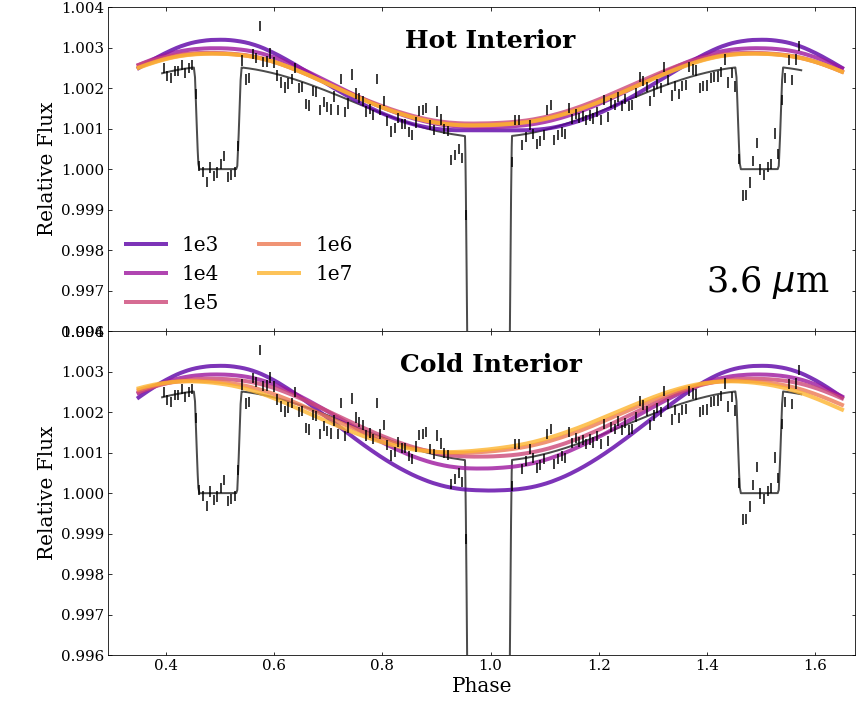}
    \includegraphics[width=0.49\textwidth]{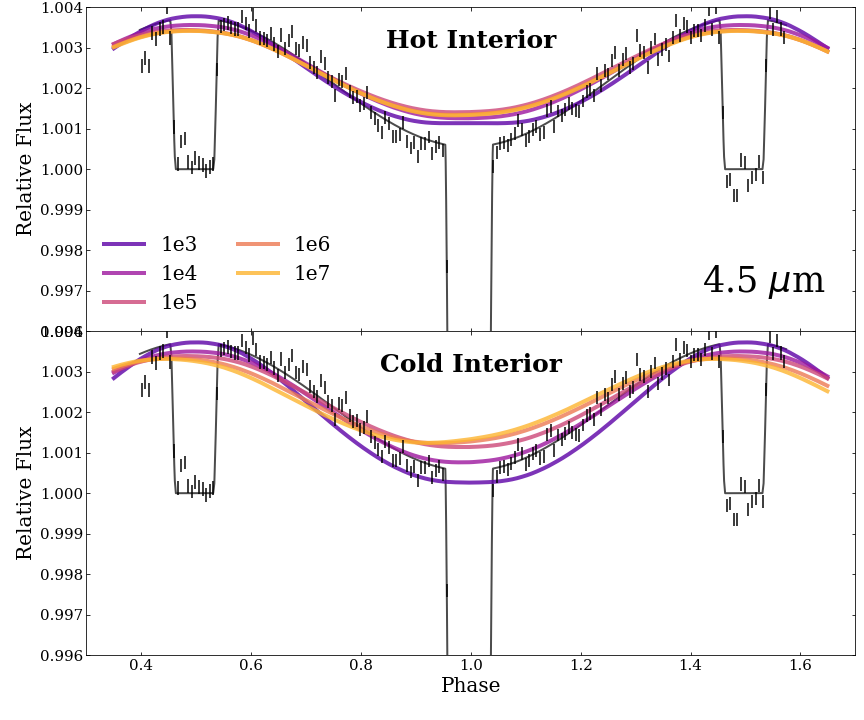}
    \caption{
    \textbf{Left:} 3.6 $\micron$ Spitzer phase curves compared to our post-processed GCM predictions. \textbf{Right:} Same for 4.5 $\micron$. In both panels we show our hot interior models on the top, with the cold interior models on the bottom. The various drag timescales considered are shown in different colors.}
    \label{fig:model_data_comp}
\end{figure*}
\begin{figure}
    \centering
    \includegraphics[width=0.49\textwidth]{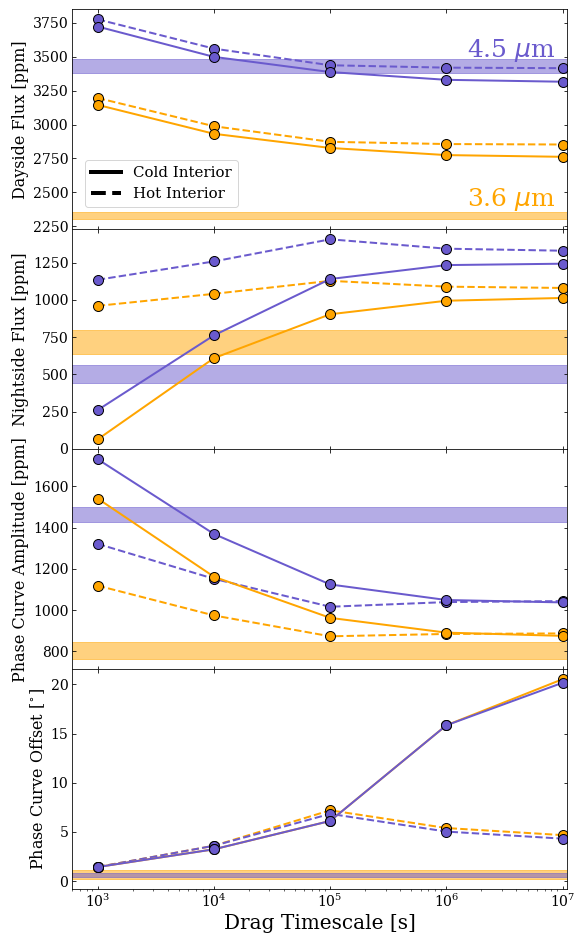}
    \caption{
    Model phase curve parameters as a function of drag timescale. From top to bottom, day side planet-to-star flux ratio, night side planet-to-star flux ratio, phase curve amplitude, and phase curve offset for the cold interior (solid lines) and hot interior (dashed line) models. 3.6 $\micron$ models and Spitzer results are shown in orange, while 4.5 $\micron$ is shown in blue. All of the properties of the $4.5~\micron$ phase curve are well matched by our cold interior models with strong frictional drag (i.e., short drag timescales). None of the models can entirely reproduce the behavior of the 3.6 micron observed phase curve -- specifically its low dayside flux and small amplitude -- which may indicate the persistence of systematics in that data set or shortcomings of our double gray GCM framework.}
    \label{fig:model_data_comp_quantities}
\end{figure}

Figure \ref{fig:model_data_comp} shows our Spitzer phase curve results compared to the hot and cold interior models discussed in Section \ref{sec:GCM}. The left panels show the 3.6 $\micron$ best fit phase curve and data compared to the hot interior models on the top, and the cold interior models on the bottom. The right panels show the same for 4.5 $\micron$. Each panel contains models with the five different drag timescales considered in our GCMs. Qualitatively, we find better agreement between the shapes of the modeled phase curves and the observations for short drag timescales (i.e.\ strong drag), particularly when considering the phase offset and night side temperature. The comparison between the data and models is explored further and in more detail below and in Figure \ref{fig:model_data_comp_quantities}. The smaller phase amplitude of our 3.6 $\micron$ phase curves, combined with the hotter night side temperature at 3.6~$\micron$, suggest that the 3.6 $\micron$ channel probes deeper into the planet's atmosphere. However, the differing eclipse depths within the previous literature, combined with the centroid jumps between AORs and the inconsistency of the two observed phase curves when fit independently, suggests that these 3.6 $\micron$ data may be unreliable. 

Figure \ref{fig:model_data_comp_quantities} compares observed phase curve parameters (day side and night side flux, phase curve amplitude, and phase curve offset) to model predictions for both the hot and cold interior cases. Our Spitzer 4.5 $\micron$ phase curve is most consistent with the cold interior models that have strong drag ($\tau_\mathrm{drag} \le 10^4~\mathrm{s}$). Models with weaker drag have both phase curve amplitudes that are smaller and phase curve offsets that are larger than found in our 4.5 $\micron$ phase curve. As a result, we rule out all weak drag models. This may point toward the presence of strong frictional drag due to Lorentz forces, which has been inferred from previous phase curve observations of ultra-hot Jupiters \citep{Kreidberg:2018aa,Arcangeli:2019aa}.
Though the day side flux at $4.5~\micron$ is well matched by a range of models, the post-processed GCMs consistently over-predict the $3.6~\micron$ day side flux. 
However, the GCM with a cold interior and $\tau_\mathrm{drag} = 10^4~\mathrm{s}$ does approximately match the observed night side flux at $3.6~\micron$. In general, the 3.6 $\micron$ data only moderately agree with the cold interior models. While this may be explained by the 3.6 $\micron$ data being unreliable, GCMs have generally struggled to match Spitzer observations \citep[e.g.][]{Stevenson2017a,Zhang2018,Crossfield2020}. Typically, 3D models are unable to capture all physics that is important due to their computational costs, and simplifying assumptions (e.g. clouds, opacity sources) that are made can bias models in ways that produce discrepant results with observations. However, because of our 4.5 $\micron$'s agreement with our models, we suggest that additional data for WASP-76b is necessary to determine the accuracy of the Spitzer 3.6 $\micron$ phase curve.   

As discussed above, and shown here, the 4.5 $\micron$ observations are well explained by the cold interior, strong drag models. Of note, this model produces a thermal inversion at pressures of approximately 1 bar to 10$^{-1}$ bar on the dayside (see Figure \ref{fig:tpgcm}). \cite{Fu:2020aa} require a thermal inversion on the dayside of WASP-76b to explain their Hubble Space Telescope eclipse spectrum, however their PHOENIX models suggest it occurs at higher in the atmosphere around 10$^{-2}$ bar. As a result, we expect that the strength, location, and presence of the inversions in our models may be impacted by the lack of clouds or enhanced metal opacity. Regardless, the Spitzer observations presented here also show evidence of a dayside temperature inversion with excess emission at 4.5 $\micron$, as well as a lack of inversion on the nightside as seen by the lower nightside emission at 4.5 $\micron$. If the observational results from the 3.6 $\micron$ Spitzer phase curve prove robust to future follow up, WASP-76b is one of the first planets with a clearly detected dayside inversion transitioning, as expected, to no measured inversion on the planet's nightside. This result follows in the footsteps of the similar ultra-hot Jupiter WASP-103b \citep{Kreidberg:2018aa}. The presence of this transition in these Spitzer data suggests that WASP-76b is an ideal target for future follow up observations.

Our requirement of strong drag to explain the 4.5 $\micron$ observations is notable in comparison to recent observations of WASP-76b at high spectral resolution. \cite{Ehrenreich:2020aa} detected a strong blue shift of the neutral atomic iron line during transit, necessitating strong day-to-night winds of $\approx 5 ~\mathrm{km}~\mathrm{s}^{-1}$ that are only achieved in the weak drag GCMs presented in this work that have $\tau_\mathrm{drag} \ge 10^5~\mathrm{s}$. Additionally, the asymmetric iron condensation signal observed by \cite{Ehrenreich:2020aa} requires a significant temperature differential between the eastern and western limbs. A large temperature contrast between the eastern and western limbs is a feature of the weak drag GCMs, while the GCMs with strong drag ($\tau_\mathrm{drag} \le 10^4~\mathrm{s}$) have a more longitudinally symmetric temperature distribution. Follow-up ESPRESSO observations by \cite{Tabernero:2020aa} also show strong blueshifts in a variety of atomic lines, including those of Mn, Na, K, and Li. 

Post-processing of the GCMs presented in this work with a ray tracing radiative transfer model adapted from that of \cite{Rauscher:2014} and \cite{Flowers:2018aa} at high resolution requires weak drag to provide the best match to these high spectral resolution observations (\citealp{Savel:2021aa}, see also \citealp{Wardenier:2021td}). One possibility is that the dynamics of the high-altitude regions probed by high-resolution spectroscopy are driven by electrodynamics, distinct from the largely neutral flow of the deeper levels probed by Spitzer observations. At high altitudes, electrons are coupled to the magnetic field and ions collisionally couple to the neutral flow, providing a large-scale ion ``drag'' force that can act to both accelerate and decelerate the flow \citep{Koskinen:2014}. The Rayleigh drag scheme applied in the GCMs presented here does not fully characterize the effect of ion drag on the flow in the upper atmospheres of ultra-hot Jupiters. As a result, the GCMs may overestimate the effect of frictional drag on the circulation at the low pressures probed by high-resolution transmission spectroscopy.


The lack of a large hotspot offset for WASP-76b is similar to previous observations of WASP-103b \citep{Kreidberg:2018aa} and WASP-18b \citep{Maxted2013} -- planets with similar temperatures to WASP-76b, and in WASP-103b's case, a comparable gravity. Small phase curve offsets may point toward the presence of frictional drag that acts to reduce the speed of the equatorial superrotating jet and has been invoked to explain observations of cooler hot Jupiters (e.g., \citealp{Knutson2012}). Additionally, they may point towards the presence of enhanced atmospheric opacity, which acts to reduce the pressure that is probed and decrease the resulting hot spot offset \citep{Lewis:2010,Kataria:2014}. Notably, GCMs for all three planets can only reasonably match the observed phase curve offset and amplitude when invoking strong atmospheric frictional drag with $\tau_\mathrm{drag} \lesssim 10^4~\mathrm{s}$ \citep{Kreidberg:2018aa,Arcangeli:2019aa}.  However, as WASP-18b has a large mass of $\approx 10~M_\mathrm{Jup}$, these three planets cover over an order of magnitude in surface gravity. The fact that comparable frictional drag timescales can provide a reasonable match to their phase curves suggests that there are similar atmospheric physics at play over a broad range of planetary mass and surface gravity. 

We find that simulations of the atmospheric circulation of WASP-76b including the hot interior expected from evolutionary calculations assuming deep heating  \citep{Thorngren:2017,Thorngren:2019aa} produce phase curve amplitudes that are too small to match the observed $4.5~\mu\mathrm{m}$ phase curve. Instead, our results are consistent with a small upward heat flux at the bottom of the atmosphere. One possible explanation for the low internal heat flux is for WASP-76b to lie on the low end of the bulk metallicity distribution for gas giants with similar mass \citep{Thorngren16}. A low bulk metallicity 
would reduce the amount of deposited heating in the interior required to explain its present-day radius \citep{Bodenheimer:2003,Fortney06,Burrows:2007bs,Fortney07,Baraffe08}, leading to a corresponding reduction in the expected internal temperature \citep{Thorngren:2017}. 
Another possible explanation for the low internal heat flux is that the atmosphere drives a downward heat flux into the interior \citep{Guillot_2002,showman_2002,Youdin_2010,Tremblin:2017}. This may offset the thermal flux from the planetary interior and further reduce the interior cooling rate. 
However, recent observations of re-inflation of hot Jupiters orbiting both main-sequence \citep{Hartman2016,Thorngren:2021aa} and post-main-sequence \citep{Grunblatt2016,Grunblatt:2017aa,Grunblatt:2019aa} stars require deposited heating within the deep interior, rather than downward heat transport by a shallow atmospheric circulation, to explain the rapid timescales of planetary re-inflation \citep{Wu:2013,Lopez:2015,Ginzburg:2015a}. 

The GCMs presented in this work did not consider night side cloud formation, which could potentially enhance the phase curve amplitude \citep{Mendonca:2018aa,Parmentier:2020aa,Roman:2020aa}. Additionally, they did not consider optical absorbers that can significantly affect the pressure of the visible photosphere and day-night heat redistribution \citep{Fortney:2008}. However, cloud formation is expected to be suppressed in the atmospheres of gas giants at ultra-hot equilibrium temperatures similar to that of WASP-76b \citep{Gao:2020aa}. The recent GCM simulations of \cite{Roman:2020aa} including radiatively active clouds found that clouds have only a minor effect on the emitted night side flux and phase curve amplitude of ultra-hot Jupiters. 
In this work, we also considered GCMs with strong drag that suppresses day-night heat transport, resulting in a cooler night side temperature than in cases with weak drag. The cooler night side temperatures in our simulations with strong drag may be more favorable for nucleation of silicate clouds \citep{Powell:2018aa,Gao:2020aa,Helling:2021aa}. Both night side cloud formation and optical absorbers reducing the visible photosphere pressure may reduce both the emergent flux from the night side and the phase curve offset \citep{Parmentier:2020aa}, both of which are over-predicted in our double-gray cloud-free GCMs with $\tau_\mathrm{drag} \ge 10^4~\mathrm{s}$. However, the hotter night sides in our simulations with a hot interior may limit the effect of clouds on the resulting phase curve, further necessitating cold interior models to match the Spitzer observations. Future work is needed to determine to what extent cloud formation may impact the observable properties of ultra-hot Jupiters with strong frictional drag and/or a high-entropy interior. 

\section{Conclusions}
\label{sec:conc}
We measured the thermal phase curve of the ultra-hot Jupiter WASP-76b with Spitzer IRAC's 3.6 and 4.5 $\micron$ channels. Below we outline our key conclusions from this observation.
\begin{enumerate}
    \item WASP-76b has an ultra-hot day side, with a day side brightness temperature of $2471 \pm 27~\mathrm{K}$ at $3.6~\mu\mathrm{m}$ and $2699 \pm 32~\mathrm{K}$ at $4.5~\mu\mathrm{m}$. The night side brightness temperature of WASP-76b at $3.6~\mu\mathrm{m}$ is $1518 \pm 61~\mathrm{K}$, much hotter than the $1259 \pm 44~\mathrm{K}$ night side temperature at $4.5~\mu\mathrm{m}$. As a result, we find that the phase curve amplitude at $3.6~\mu\mathrm{m}$ of $804.0 \pm 42.5~\mathrm{ppm}$ is significantly lower than the $1464 \pm 38~\mathrm{ppm}$ amplitude at $4.5~\mu\mathrm{m}$. 
    \item We observe a small phase curve offset at $3.6~\mu\mathrm{m}$ of $0.68 \pm 0.48^{\circ}$ and a similarly small phase curve offset of $0.67 \pm 0.2^{\circ}$ at $4.5~\mu\mathrm{m}$. These small observed phase curve offsets are similar to recent phase curves of the ultra-hot Jupiters WASP-18b and WASP-103b that receive a comparable level of incident flux to WASP-76b. In agreement with previous WASP-76b observations, this is suggestive of a thermal inversion on the planet's dayside transitioning to a lack of inversion on the planet's nightside. We caution that the 3.6 $\micron$ data is less reliable than the 4.5 $\micron$ data and follow up observations are necessary to confirm the presence of the dayside inversion.
    \item We calculated a grid of GCMs for the atmospheric circulation of WASP-76b including the impact of large-scale drag on the flow, parameterizing the effects of magnetism and/or turbulence. We find that the small observed phase curve offset at both $3.6~\mu\mathrm{m}$ and $4.5~\mu\mathrm{m}$ and the large phase curve amplitude at $4.5~\mu\mathrm{m}$ require strong frictional drag to explain. We find that a drag timescale of $\lesssim 10^4~\mathrm{s}$ is required to match the observed $4.5~\micron$ phase curve, which is similar to the level of drag required from previous GCM simulations of WASP-18b and WASP-103b.
   However, we note that we cannot explain the relatively small phase curve amplitude observed at $3.6~\mu\mathrm{m}$ with any GCM in our model suite. Of note, though, is the relatively large impact of systematics in these 3.6 $\micron$ data, potentially pointing to unreliability of the 3.6 $\micron$ Spitzer observations of WASP-76b. One source of this may be the companion star that is unresolved from WASP-76 in these data at a separation of only 0.3 pixels, exacerbating the strong 3.6 $\micron$ intrapixel effect.
    \item Due to the inflated radius of WASP-76b, we also consider atmospheres that overlie a high-entropy interior. We find that cloud-free, double-gray GCMs without a high-entropy interior provide a better match to the observed night side flux and phase curve amplitude at $4.5~\mu\mathrm{m}$. This may imply that the interior evolution of WASP-76b does not significantly affect its atmospheric circulation at the photosphere, or that clouds strongly reduce the outgoing thermal emission from the night side of the planet. 
    
\end{enumerate}

\software{\\ IPython \citep{ipython},
\\ Matplotlib \citep{matplotlib},
\\ NumPy \citep{numpy,numpynew},}

\acknowledgments
E.M.M.\ acknowledges support from JHU APL's Independent Research And Development program. T.D.K.\ acknowledges funding from the 51 Pegasi b Fellowship in Planetary Astronomy sponsored by the Heising–Simons Foundation. K.B.S.\ and J.L.B.\ acknowledge support for this work from NASA through awards issued by JPL/Caltech (Spitzer programs 13038 and 14059). M.\ Mansfield acknowledges funding from a NASA FINESST grant. E.M.-R.K. acknowledges support from the NASA Astrophysics Theory Program (grant NNX17AG25G) and the
Heising-Simons Foundation. J.M.D. acknowledges funding from the European Research Council (ERC) under the European Union's Horizon 2020 research and innovation programme (grant agreement no. 679633; Exo-Atmos). J.M.D acknowledges support by the Amsterdam Academic Alliance (AAA) Program. This work was completed with resources provided by the University of Chicago Research Computing Center.

\appendix

\begin{deluxetable}{l c c c}
    \tablecolumns{4}
    \tabletypesize{\footnotesize}
    \tablecaption{Best fit parameters}
    \label{table:all_params}
    \tablehead{
        \colhead{Label} &
        \colhead{wa076bo21}  &
        \colhead{wa076bo11} &
        \colhead{wa076bo12}  
        }
    \startdata
Transit Midpoint [bjdtdb] & 1.81968 & 0.91826 & 1.8451  \\
Midpoint Offset [bjdtdb] & 2457858.0 & 2457877.0 & 2458229.0 \\
Rp/Rs & 0.106 & 0.10048 & 0.10048  \\
Per. [days] & 1.8099 & 1.8099 & 1.8099  \\
a/Rs & 4.07463 & 4.0476 & 4.0476  \\
a [au] & 0.033 & 0.033 & 0.033  \\
cos(i) & 0.0349 & 0.0349 & 0.0349  \\
Limb darkening, u1 & 0.081 & 0.08 & 0.08  \\
Limb darkening, u2 & 0.097 & 0.13 & 0.13  \\
T$_s$ & 6250.0 & 6250.0 & 6250.0  \\
lo0 [deg] & 0.67 & 0.68 & 0.68  \\
sph0 ($Y_{0}^{0}$) & 0.00063 & 0.00048 & 0.00048  \\
sph3 ($Y_{1}^{1}$) & 0.00039 & 0.00022 & 0.00022  \\
Ramp, Quad. Term & -- & -0.00111 & -- \\
Ramp, Lin. Term & -0.00284 & 0.00296 & 0.00215  \\
Ramp, Constant & 1.0 & 1.0 & 1.0  \\
Ramp, Offset [bjdtdb] & 1.8 & 0.8 & 1.8  \\
Constant & 83526.0 & 130709.9 & 127708.2  \\
PRF Width, Lin. Term in x & -- & -0.02976 & 0.12086 \\
PRF Width, Quad. Term in x & -- & 0.02154 & -0.17816 \\
PRF Width, Lin. Term in y & -- & -0.08252 & 0.03993 \\
PRF Width, Quad. Term in y & -- & 0.49359 & -0.37031 \\
PRF Width, Constant Term & -- & 1.0 & 1.0 \\
PRF Width, Offset Term & -- & 0.5 & 0.5 \\
    \enddata
    \tablecomments{We use the lon0/lat0 options in \texttt{spiderman}, not the corresponding spherical harmonics terms.}
\end{deluxetable}

\bibliography{references_paste}





\end{document}